\documentclass{article} % For LaTeX2e
\usepackage{iclr2026_conference,times}
\usepackage{amsmath,amsfonts,bm}

\def\eqref#1{equation~\ref{#1}}
\def\1{\bm{1}}

\DeclareMathAlphabet{\mathsfit}{\encodingdefault}{\sfdefault}{m}{sl}
\SetMathAlphabet{\mathsfit}{bold}{\encodingdefault}{\sfdefault}{bx}{n}

\usepackage[utf8]{inputenc} % allow utf-8 input
\usepackage[T1]{fontenc}    % use 8-bit T1 fonts
\usepackage{hyperref}       % hyperlinks
\usepackage{url}            % simple URL typesetting
\usepackage{booktabs}       % professional-quality tables
\usepackage{amsfonts}       % blackboard math symbols
\usepackage{amsmath}       % blackboard math symbols
\usepackage{nicefrac}       % compact symbols for 1/2, etc.
\usepackage{microtype}      % microtypography
\usepackage{graphicx}
\usepackage{xcolor}         % colors
\usepackage{subcaption}
\usepackage{caption}
\usepackage{array}
\usepackage{pifont}
\usepackage[inkscapeversion=1.3]{svg}

\svgsetup{inkscapelatex=false}

\title{Emergence of Spatial Representation in an Actor-Critic Agent with Hippocampus-Inspired Sequence Generator}

\author{%
  Xiao-Xiong Lin\thanks{  Correspondence to: \texttt{ln.xxng@gmail.com}}
  \quad Yuk-Hoi Yiu\quad Christian Leibold \\
  \\
  Faculty of Biology \& Bernstein Center Freiburg \&
  BrainLinks-BrainTools Center\\
  University of Freiburg, Freiburg, Germany\\
}
\iclrfinalcopy % Uncomment for camera-ready version, but NOT for submission.
\begin{document}

\maketitle

\begin{abstract}

Sequential firing of hippocampal place cells is often attributed to sequential sensory drive along a trajectory, and has also been attributed to planning and other cognitive functions. Here, we propose a mechanistic and parsimonious interpretation to complement these ideas: hippocampal sequences arise from intrinsic recurrent circuitry that propagates transient input over long horizons, acting as a temporal memory buffer that is especially useful when reliable sensory evidence is sparse.
We implement this idea with a minimal sequence generator inspired by neurobiology and pair it with an actor–critic learner for egocentric visual navigation. Our agent reliably solves a continuous maze without explicit geometric cues, with performance depending on the length of the recurrent sequence. Crucially, the model outperforms LSTM cores under sparse input conditions (16 channels, $\sim2.5\%$ activity), but not under dense input, revealing a strong interaction between representational sparsity and memory architecture.
Through learning, units develop localized place fields, distance-dependent spatial kernels, and task-dependent remapping, while inputs to the sequence generator orthogonalize and spatial information increases across layers. These phenomena align with neurobiological data and are causal to performance. Together, our results show that sparse input synergizes with sequence-generating dynamics, providing both a mechanistic account of place cell sequences in the mammalian hippocampus and a simple inductive bias for reinforcement learning based on sparse egocentric inputs in navigation tasks.
Code: \url{https://github.com/xiaoxionglin/SF_hipposlam}

%%% add control: training just the visual encoder; HiPPO CA3

\end{abstract}

\section{Introduction}

% \begin{figure}
%     \centering
%     \includegraphics[width=1.\linewidth]{figs/theory_rnn.pdf}
%     \caption{Theta sequences.
%     \textbf{A} Illustration of theta sequences observed in rodent hippocampus. In each theta cycle, R=3 neurons are activated and the activation propagates in a sequence of $\ell$=5 neurons over L=3 theta cycles (cf. eq.~\ref{eq:S-and-J}).
% \textbf{B} Current baseline understanding of the theta sequences, with unspecified recurrent connections and sequential inputs driving sequential activation of place cells.
% \textbf{C} Intrinsic theta sequence hypothesis, a parsimonious account where the recurrent connections support generating long horizon sequential activity without sequential external inputs. }
%     \label{fig:fig1_hypo}
% \end{figure}

Hippocampal place cells track the animal’s location during navigation \citep{OKeefe1971-mr} and they fire in sequence reflecting the behavioral order of the place fields~\citep{Foster2007-wz}. Spatial locations are thereby thought to serve as anchors for episodic memories~\citep{aronowitz2023space}, a view reinforced by observations of “look-ahead” sequence replay linked to trajectory planning~\citep{Foster2007-wz,Kay2020-xz}. 
On the other hand, hippocampal neurons are also found to fire at successive moments not explainable by location alone~\citep{eichenbaum2014time}, suggesting the place cell sequences could reflect timing rather than spatial input.

% The hippocampus is an essential brain region for episodic memory and spatial navigation. Damage impairs episodic memory formation in humans \citep{Scoville2000-jm}, and rodent recordings revealed place cells tuned to specific locations \citep{OKeefe1971-mr}. Spatial locations are thus thought to serve as anchors for episodic memories~\cite{aronowitz2023space}, a view reinforced by observations of “look-ahead” replay sequences linked to trajectory planning~\cite{Foster2007-wz,Kay2020-xz}.

The intertwined spatial and temporal representations in hippocampus has been touched upon in many recent computational models. Successor representations interpret place cells as predictive states \citep{stachenfeld2017hippocampus,Mattar2018-px}; reservoir models emphasize pre-existing dynamics in shaping place cell sequences \citep{leibold_model_2020}; probabilistic approaches model place cells as latent states inferred from successive inputs \citep{raju_space_2024}; and self-supervised methods refine spatial tuning by exploiting temporal smoothness of the trajectory~\citep{wang_time_2024}. While these approaches reproduce place-like activity and even sequential patterns, they rarely address explicitly where hippocampal sequences originate. Many bottom-up studies instead emphasize that structured representations emerge from circuit dynamics and physiological constraints, suggesting mechanistic models are needed to link anatomy, dynamics, and representational patterns~\citep{lin2023,buzsaki2018sequence_generator,schaeffer2022nofreelunch}.

\begin{figure}
    \centering
    \includegraphics[width=1.\linewidth]{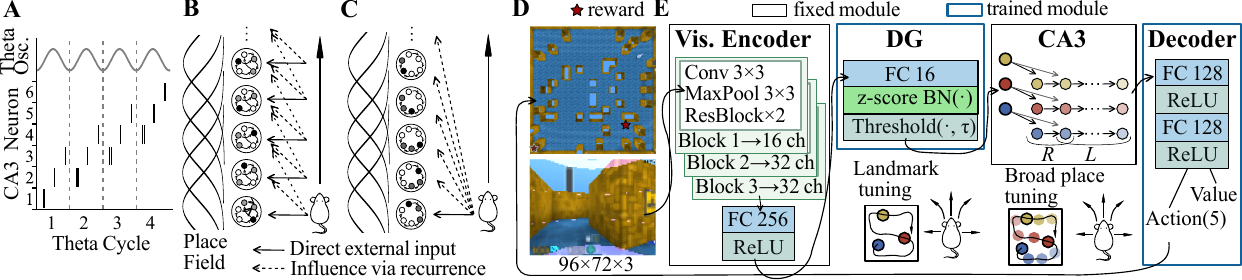}
    \caption{
        Model summary.
        \textbf{A}: Illustration of theta sequences observed in rodent hippocampus. In each theta cycle, $R=3$ neurons are activated and the activation propagates over $L=4$ theta cycles in a sequence of $\ell=L+R-1=6$ neurons.
        \textbf{B}: The theta sequences are thought to be driven by sequential inputs despite the recurrent connections in hippocampus. 
        %Left: spatial tunings of the sequentially activated cells.
        \textbf{C}: The recurrent connections could support generating long horizon sequential activity without sequential external inputs. 
        \textbf{D}: A virtual environment (\(19\times19\) tiles) was constructed using DeepMind Lab with walls randomly placed on 15~\% of the tiles. Wall layouts are kept fixed across episodes,  with an invisible reward near the bottom right. 
        \textbf{E}: 
        The agent receives a first person perspective visual input that is processed via an encoder (shallow ResNet with 3 convolutional blocks; matching the SOTA in DeepMind Lab environment \citep{espeholt2018impala}, pretrained and fixed in our experiments).
        These outputs are linearly mapped to $F=16$ features (FC: fully connected layer), and then sparsified using batch normalization and high thresholding (\(\tau=2.43\)), such that the percentage of activation (\(\sim2.5\%\)) matches the sparse activity of DG granular cells that project to CA3.
        CA3 is modeled via sequences of neuron activations, each sequence evoked by a distinct DG input feature. The activity of all CA3 neurons are then flattened and linearly mapped to the decoder multilayer perceptron.
        CA3 is hard coded to isolate the effect of long range integration. The DG and decoder modules are trained. 
}
    \label{fig:fig1_model}
\vspace{-8pt}
\end{figure}

We propose a parsimonious account directly addressing this question: hippocampal sequences arise from intrinsic recurrent circuitry in CA3 that can propagate activity over long timescales in the absence of input (Fig.~\ref{fig:fig1_model}A-C). The CA3 sequence generator receives sparse inputs from dentate gyrus (DG) to yield localized spatial codes in CA3 that support navigation \citep{leibold_model_2020,leibold2022neuralkernel}. The sparse-input regime is not incidental but central: it reflects the ecological reality that navigation is often guided by only a few reliable landmarks amid abundant sensory noise; the biology of DG granule cells, which fire at extremely low rates; and the computational advantages of high-capacity, low-interference codes that promote compositionality and generalization.

This mechanism mirrors recent key ideas from machine learning. State-space models and structured linear RNNs preserve long-range information by expanding inputs into a high-dimensional temporal feature space before compressing them via shallow nonlinear readouts \citep{fu2022hungry,gu2020hippo,gu2021efficiently}. Our model resonates with this principle: DG sparsification provides a low-activity code that is sustained and expanded by intrinsic recurrence, offering a rich set of features for downstream policy learning.

To test this hypothesis, we implement an agent with a DG-like sparsification module, a recurrent sequence generator (CA3 proxy), and an actor–critic learner for egocentric navigation. We show that sequence generation and sparse input synergize, outperforming LSTMs of comparable size in the sparse-input regime, while LSTMs remain competitive under dense input. Moreover, place fields, DG orthogonalization, and task-dependent remapping emerge naturally during training.

These results suggest that a sequence-based reservoir, inspired by CA3, is well suited for constructing spatial representations from sparse low-bandwidth inputs. The synergy between sparse coding and intrinsic sequence dynamics thus offers both a mechanistic explanation for hippocampal sequences and a simple inductive bias for reinforcement learning in navigation tasks.

\section{Methods}

\subsection{Virtual Environment}
%the game is 60 fps, each action is repeated for 8 frames until the next observation
%I'll change the x-axis to "step"/"action" instead of "frame"
We consider an agent navigating a continuous environment with sparse obstacles and uniform visual textures. The environment was simulated using DeepMind Lab \citep{beattie2016deepmindlab} (Figure~\ref{fig:fig1_model}D). The wall layouts were kept fixed for repeated episodes unless otherwise stated. The environment was designed such that spatial relations between locations cannot be trivially inferred from the similarity of their corresponding visual features and that it allows multiple possible trajectories from a given location to the target.

In each episode, the agent was initially placed at a random location at least 5 units from the goal. An episode is finished when the agent reaches the goal or a maximum of 7200 frames (with action repeat=8 corresponding to a maximum of 900 action steps, i.e. time steps in the model). For the map displayed in Fig.~\ref{fig:fig1_model}D, the reward was placed near the bottom right corner and then moved to the lower left corner to test generalization. 
%POTENTIAL_DELETE
%Since our navigation task does not require complex actions, we reduced the action space used in IMPALA~\citep{espeholt2018impala}, to facilitate training speed (Table~\ref{tab:reduced_action_set}).

\subsection{Visual processing}
The visual encoder is intended to mimic visual cortical preprocessing, extracting general-purpose visual features for the hippocampus. We pretrained a ResNet encoder \citep{he2015deepresiduallearningimage_resnet} in combination with an LSTM core as in \citet{espeholt2018impala} and kept it fixed when training our hippocampus model. We also tested a variant using the layer 2 output from ResNet pretrained on Imagenet. The model produces similar performance (Fig.~\ref{fig:training_layer2}).

\subsection{Hippocampus Model}

\paragraph{Dentate gyrus as a sparsification module.}
A main cortical input to the hippocampus is routed through the dentate gyrus (DG) \citep{Amaral}. DG activity is characteristically sparse, with only about $2$–$5\%$ of granule cells active in a given environment \citep{Henze2000-ag,leutgeb_pattern_2007}. We model the DG as a linear projection of visual features followed by batch normalization (running estimates of mean and variance, retention rate $0.95$ per minibatch) and a high activation threshold to match the sparsity of activity ($\sim2.5\%$). 

\paragraph{CA3 as a sequence-generating shift register.}
We model CA3 as a linear RNN shift register that propagates inputs as theta sequences (Fig.~\ref{fig:fig1_model}C and E, CA3) and, for this paper, we keep it fixed to isolate the effects of intrinsic sequence generation from effects potentially induced by recurrent plasticity.
The DG provides input $u_t \in \mathbb{R}^F$ over $F$ features. Each feature is assigned a dedicated prewired sequence of
length $\ell$, so the CA3 state is $X_t \in \mathbb{R}^{F\ell}$. Motivated by hippocampal theta sequences
\citep{Dragoi2006-qq,Foster2007-wz,leibold_model_2020}, we parameterize the total number of sequence units as
$\ell = L + (R-1)$, where $L$ is the number of theta cycles and $R$ the number of active units per cycle.

\paragraph{Single-feature dynamics ($F=1$).}
Let $x_t \in \mathbb{R}^{\ell}$ denote the CA3 state for a single feature and $u_t \in \mathbb{R}$ the corresponding DG input.
Sequence propagation is
\begin{equation}
x_{t+1} \;=\; S\,x_t \;+\; J\,u_t,
\label{eq:single-feature}
\end{equation}
where $S \in \mathbb{R}^{\ell \times \ell}$ is the shift operator and $J \in \mathbb{R}^{\ell \times 1}$ injects
the input into the first $R$ slots:
\begin{equation}
S =
\begin{bmatrix}
0 & 0 & \dots & 0\\
1 & 0 & \dots & 0\\
 & \ddots & \ddots & \vdots\\
0 & \dots & 1 & 0
\end{bmatrix}_{\ell\times\ell},
\qquad
J = \bigl[\underbrace{1,1,\dots,1}_{R\text{ times}}\;,\underbrace{0,0,\dots,0}_{L-1\text{ times}}\bigr]^T \in \mathbb R^{\ell\times1}.
\label{eq:S-and-J}
\end{equation}
Thus a transient input $u_t$ creates activity in the first $R$ positions which is then shifted one step per timestep
along the length-$\ell$ register.

\paragraph{Multiple features ($F>1$).}
Each feature dimension evolves independently under the same dynamics. Stacking all $F$ sequences with Kronecker expansion gives
the block-structured update
\begin{equation}
A \;=\; I_F \otimes S, \qquad B \;=\; I_F \otimes J,
\label{eq:A-and-B}
\end{equation}
where $I_F$ is an identity matrix that has the size of the number of DG features. $S$ and $J$ are the same as the single feature case (For an extended explanation, see Appendix~\ref{app:mf}).
and the full CA3 dynamics
\begin{equation}
X_{t+1} \;=\; A\,X_t \;+\; B\,u_t,
\label{eq:CA3-full}
\end{equation}
with fixed recurrent matrix $A$ and input matrix $B$.

\subsection{Actor Critic Network}

The activity of hippocampal units is fed into a decoder module with two linear/fully connected (FC) layers with ReLU activation. Actions and value are computed as the linear readout from the decoder module. 
We trained the agent with a standard advantage actor–critic objective (policy-gradient + value-baseline + entropy regularization) as implemented in Sample Factory \citep{petrenko2020sf, espeholt2018impala}.
%We freeze the visual encoder and fix CA3 dynamics to isolate the contribution of sequence propagation; only the DG projection and decoder are optimized by RL.”

%We used asynchronous training
%We used the package Sample Factory \citep{petrenko2020sf} to train our %agent. It enables training 
% *********************
% *PPPPPPPPPPPPPPPPPPP*
% *PPPPP*PPPPPPP*PPP*P*
% *PPPPPPPPPPPPPPP*PPP*
% *PPP*PPPPPPPPPP*PPPP*
% *P*PPPPPPPPPPPPPPPPP*
% *PPPPPPPPPPP**PPPPP**
% *PPPPP*P*PPP*PPPPPPP*
% *PPPPP*PPPPPPP*PPP*P*
% *PPP*PPPPP*PPPPPPPPP*
% *PP*PPPPPPPPP*PPPP*P*
% *PPPPPPPPP*PPPPPPPPP*
% *PPP*PPPPP*PP * P *P*
% *P*PPPPPPPPP*      P*
% *PP**PPP*P**   **   *
% *PPPPPPPPP     G    *
% **PPPPPPPPP *    *  *
% *PPPPPPPPPPP       P*
% *P*PPPPPPPP*P  * *PP*
% *PPPPPPP*PPPPP P PPP*
% *********************

\begin{figure}
    \centering
    \includegraphics[width=1\linewidth]{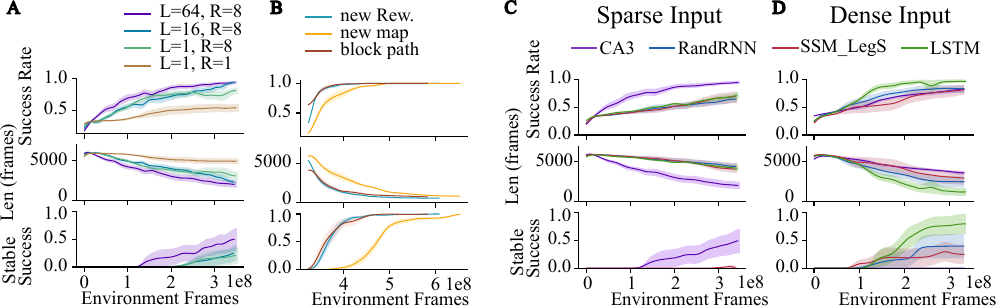}
    \caption{
    Training performance
    \textbf{A} Agents with different sequence length $L$ and repeat $R$. Line and shaded area are mean and s.e.m. across 6 random seeds. Success Rate: fraction of episodes where the goal is reached. Len: number of frames to reach the goal. Stable success: the rate of having 100 consecutive successful episodes. Metrics were Gaussian smoothed, \(\sigma=6\times10^6\) frames.
    \textbf{B}
    The best performing agent with $L=64$ and $R=8$ across all seeds was tested for its transfer learning. New Rew.: new reward location at the lower left corner. new map: a randomly generated map with the same statistics. block path: new walls are added to block paths to the reward, while the previous blocking walls were removed.
    \textbf{C} and     \textbf{D}
    Performance of agents with different recurrent modules and inputs. \textbf{C}: Sparse input. \textbf{D}: Dense input where batch-normalization and high thresholding was removed. CA3: our CA3 model with $L=64$ and $R=8$. RandRNN: randomly initialized fixed RNN of the same state size. SSM\_LegS: fixed SSM HiPPO-LegS from \citet{gu2020hippo} with the same state size. LSTM: trainable LSTM with matching number of parameters.
    }
    \label{fig:training_DS}
\end{figure}
% \begin{figure}
%     \centering
%     \includegraphics[width=1\linewidth]{Neuripsfigs/ICLR/HippoL_HippoR_transfer.pdf}
%     \caption{
%     \textbf{A} Training performance of agents with different sequence length L and repeat R. Line and shaded area are mean and s.e.m. across 6 random seeds. Len: number of frames to reach the goal location. Stable success: the rate of having 100 consecutive successful episodes. Metrics were Gaussian smoothed, \(\sigma=6\times10^6\) frames.
%     \textbf{B}
%     The best performing agent with L=64 and R=8 across all seeds was tested for its transfer learning. New Rew.: new reward location at the lower left corner. new map: a randomly generated map with the same statistics. block path: new walls are added to block paths to the reward, while the previous blocking walls are removed.
%     }
%     \label{fig:training_transfer}
% \end{figure}

% \begin{figure}
%     \centering
%     \includegraphics[width=0.84\linewidth]{Neuripsfigs/ICLR/DS_Archi_HiPPO.pdf}
%     \caption{
%     Performance of agents with different recurrent modules and inputs. Left: sparse input. Right: dense input where batch-normalization and high thresholding was removed. CA3: our CA3 model with L=64 and R=8. RandRNN: randomly initialized fixed RNN of the same state size. SSM\_LegS: fixed SSM HiPPO-LegS from \citet{gu2020hippo} with the same state size. LSTM: trainable LSTM with matching number of parameters.
%     }
%     \label{fig:training_DS}
% \end{figure}

\section{Results}
\subsection{Behavioral Performance}
Training the naive agent ($L=64, R=8$) on one fixed reward location in the maze exhibits robust performance measures after about 350 million frames (Fig.~\ref{fig:training_DS}A). Thus our model effectively maintains information about sparse inputs in its trajectories of the RNN-like dynamics across time like in a reservoir~\citep{Jaeger2004-le}. This view is further supported by our simulations with reduced sequence length showing inferior performance (Fig.~\ref{fig:training_DS}A). In the most extreme case ($L=1,R=1$) without sequences, essentially a pure feedforward architecture corresponding to a brain with DG output bypassing CA3, the agent did not achieve robust behavior.  While agents with reduced sequence lengths ($L=1,16,R=8$) still can express reasonable success rates and trajectory lengths, their behavior is considerably more unstable as evidenced by the low fraction of consecutive successes (Fig.~\ref{fig:training_DS}A). 
The agent's performance is stable within a wide range of $R$ and becomes more sensitive to $R$ when running speed is slower (Fig.~\ref{fig:Hippo_R} and Tab.~\ref{tab:Hippo_R}).
Transfer learning for new reward location and blocked paths, however, requires only about 50 million frames, and transfer learning to a new map requires about 150 million frames. These indicate that the agent was able to form a generalizing representation of the map and task (Fig.~\ref{fig:training_DS}B).

The agent equipped with a sequence-generating module (CA3; eq.\ref{eq:CA3-full}) learns faster under sparse DG input than an agent in which CA3 is replaced by an LSTM core, resembling the SOTA architecture of \citet{hessel2019multi_popart} on the DMLab-30 benchmark \citep{beattie2016deepmindlab}, from which our environment is adapted. Our CA3 module also outperforms state-space model HiPPO-LegS \citep{gu2020hippo} and, as an additional control, randomly initialized RNNs, indicating that its theta-sequence dynamics provide a distinct advantage (Fig.~\ref{fig:training_DS}C).

Crucially, this advantage is confined to the sparse-input regime. Under dense input, LSTMs perform better—consistent with prior reports \citep{hessel2019multi_popart}—and our CA3 module performs worse than HiPPO-LegS and random RNNs (Fig.~\ref{fig:training_DS}D). We also tested other variants from \citet{gu2020hippo} and they performed similar to HiPPO-LegS (Tab.~\ref{tab:results_dense_sparse}, Fig.~\ref{fig:training_SSM}). These results highlight a specific synergy between sparse representations and intrinsic sequence generator.

\begin{figure}
    \centering
    \includegraphics[width=1\linewidth]{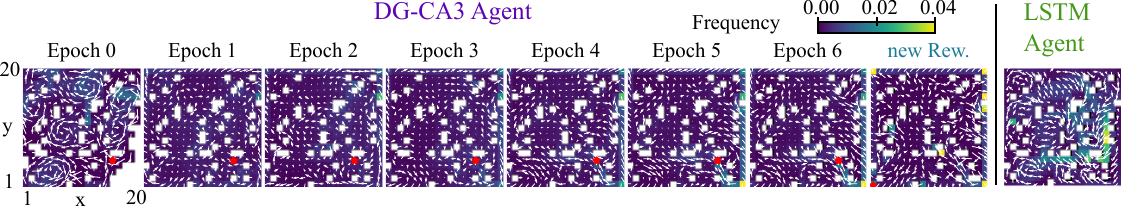}
    \caption{Evolution of occupancy over the course of learning. Color represents a normalized measure of the amount of time points the agent spent at a location. Mean head directions at a location were visualized in arrows. From epoch 5 onwards, the agent has a preference to reach the bottom right corner (independent of the random starting location) and proceed to the goal (red star) from there. The last panel shows the behavior of fully trained dense LSTM agent.}
    \label{fig:occu}
\end{figure}

\subsection{Occupancy}
We divided the training into 6 epochs and evaluated the behavior and learned representations at these checkpoints. The agent develops a stable trajectory after 4 to 5 epochs. Between Epoch 4 and Epoch 5, the agent learned to get around the obstacle in the upper part of the right edge and started to spend more time at the lower right corner before reaching the reward site (Fig.~\ref{fig:occu}). 
The agent appears to develop a strategy of visiting locations with salient input/landmarks and converging from different starting locations to the habituated paths, even after switching to new reward location. This is consistent with the typical strategies employed by animals in familiar enviroments~\citep{gibson2013head}.
In comparison, the LSTM agent under dense sensory input has fewer converging trajectories before reaching the reward (Fig.~\ref{fig:occu} and Fig.~\ref{fig:lstm-occupancy}, likely due to the goal information being more readily available in the visual input, thereby implementing a strategy more related to visual search.

\begin{table}[t]
\centering
\caption{Performance comparison across architectures. 
Training steps indicate the number of environment frames (in millions) required to reach 80\% success . ``\ding{55}'' indicates that the threshold was not reached in all random seeds within 350M frames.}
\label{tab:results_dense_sparse}
\resizebox{\linewidth}{!}{%
\begin{tabular}{lcccccc}
\toprule
Input & CA3 & Random RNN & HiPPO-LegS & HiPPO-LegT & HiPPO-LagT & LSTM \\
\midrule
Sparse (Steps)  & 173.6$\pm$77.6 & \ding{55} & \ding{55} & \ding{55} & \ding{55} & \ding{55} \\
Sparse (Succ.)  & \textbf{0.86$\pm$0.10} & 0.51$\pm$0.12 & 0.52$\pm$0.11 & 0.57$\pm$0.15 & 0.65$\pm$0.10 & 0.56$\pm$0.06 \\
Dense (Steps)   & \ding{55} & \ding{55} & \ding{55} & \ding{55} & \ding{55} & 135.9$\pm$27.6 \\
Dense (Succ.)   & 0.71$\pm$0.07 & 0.78$\pm$0.15 & 0.64$\pm$0.21 & 0.65$\pm$0.28 & 0.38$\pm$0.02 & \textbf{0.93$\pm$0.09} \\
\bottomrule
\end{tabular}%
}
\end{table}

\begin{figure}
    \centering
    \includegraphics[width=1.\linewidth]{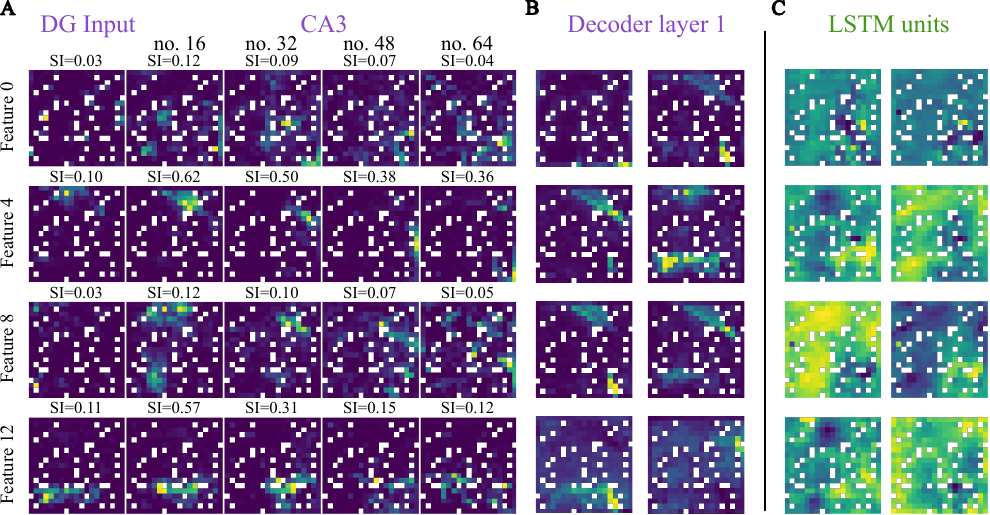}
    \caption{\textbf{A} Spatial tuning of DG and CA3 units from epoch 6. Pixel coordinates correspond to environment (black crosses correspond to walls). Each row shows the CA3 units ordered by their positions in the activity sequence. We selected 4 out of the 16 feature sequences for visualization. Spatial Information (SI): bits per time step. \textbf{B} Spatial tuning of randomly selected Decoder layer 1 units from epoch 6. \textbf{C} Spatial tuning of example LSTM output units in the dense LSTM agent. Colormaps are min-max scaled.}
    \label{fig:Place_field_seq0}
\end{figure}

\begin{figure}
    \centering
    \includegraphics[width=1\linewidth]{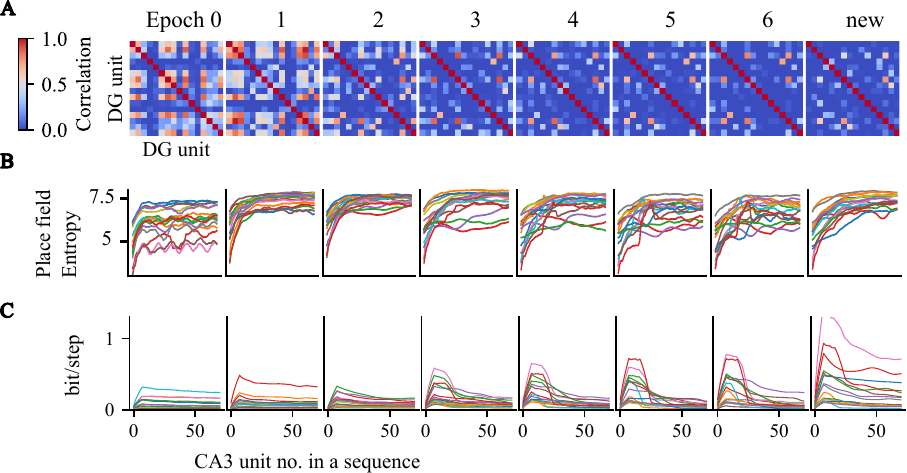}
    \caption{\textbf{A} Correlation (color code) of the 16 DG activity maps evoked by the visual features. Throughout training, the DG input to CA3 became increasingly orthogonal. 
    \textbf{B} The spread of CA3 place field measured by entropy. The units are ordered by their positions in a sequence. Each sequence is indicated by a unique color across epochs.
    \textbf{C} Same as \textbf{B} but for the spatial information of CA3 units.}
    \label{fig:Epoch_DG_entropy_SI}
\end{figure}

\subsection{Place Field Analysis}
In order to understand how the agent executes a successful strategy and whether the representations in the hippocampus-inspired model would also align with the known physiology of place cells, we further computed activity maps of the units for the different layers of the model network. For the input units (DG), we (by construction) observe sparse activation that is distributed across few locations in the map that is further refined in the subsequent linear readout in decoder layer 1 (Fig.~\ref{fig:Place_field_seq0}AB). Hidden units in the LSTM agent show non-localized tuning, distinct from typical place cells (Fig.~\ref{fig:Place_field_seq0}C).
More interestingly, learning the weights from the visual encoder to DG leads to orthogonalization, such that DG population activity over time develops a unique representation of individual locations (Fig.~\ref{fig:Epoch_DG_entropy_SI}A). 
Units farther away from DG input in CA3 sequences show broader spatial tuning (Fig.~\ref{fig:Place_field_seq0}A; quantified in Fig.~\ref{fig:Epoch_DG_entropy_SI}B) matching previous reports~\citep{Jung1993-ea,parra2021sequence_speed_width}. The place fields of DG and CA3 units gradually stabilize through learning and exhibit remapping after adapting to a new reward location, measured by shifts of the place field center of mass  (Fig.~\ref{fig:CoM}).
In contrast, the input to LSTM agents did not show orthogonalization of the learned input projection (Fig.~\ref{fig:lstm-dg-pfs}).

\subsection{Spatial Information Analysis}

Qualitative inspection of the spatial tunings in Fig.~\ref{fig:Place_field_seq0} is also quantitatively supported by the distributions of spatial information (SI)  of the neural activity maps: CA3 units activated $\sim 16$ steps after the input have larger SI (Fig.~\ref{fig:Epoch_DG_entropy_SI}C). This corroborates our training results that agents with sequence length \(L=16\) are able to acquire the task reasonably well (Fig.~\ref{fig:training_DS}). Units activated later in a sequence exhibit smaller spatial information (Fig.~\ref{fig:Epoch_DG_entropy_SI}C).
 
%SI measured by unit activity (bit/step normalized by mean activity level) %however, has an almost monotonically decreasing trend along the units in a %sequence. This suggests that the CA3 units facilitate spatial %representation by increasing the activity level compared to the sparse %input (firing at more steps due to repetition number \(R\)), besides the %goal-directed behavior converging trajectories to 

% want to remove this
Tracking the distribution of SI rates across training epochs, we observe a growing long tail indicative for the development of a neural space representation throughout all layers (Fig.~\ref{fig:SI}). Furthermore, SI also increases with increasing hierarchical level of the network layer until the first layer of the decoder network. 

To test whether SI is causal to behavior, we selectively permute the output weights of Decoder layer 1 units. When the permutation was done on the 32 units with the lowest SI, the performance was not impaired with respect to success rate and trajectory length (100\% and 1065 frames). When the permutation was done on the 32 units with the highest SI, the success rate dropped by 4.9\% and the average trajectory length increased from 1065 frame to 2794 frames.

% To better understand how the decoder layers extract spatial information from the sequence network, we visualized the effective synaptic matrices from the sequence network to the decoder layer (Fig.~\ref{fig:weights_mlp0_seqall}). The matrices reveal that decoder units generally rely multiple CA3 sequences. Thus spatial representation underlying successful goal finding is extracted from a combination of input features and their temporal history.

% \begin{figure}
%     \centering
%     \includegraphics[width=1\linewidth]{Neuripsfigs/weight_mlp0_seqs.pdf}
%     \includegraphics[width=1\linewidth]{Neuripsfigs/weight_mlp2_seqs.pdf}
%     \caption{Synaptic weights (absolute values color coded) from the sequence layer (CA3) to the first layer of the decoder network (top) and the effective connectivity (bottom)  from CA3 to the second decoder layer. The absolute values are then averaged across CA3 units in the same sequence.}
%     \label{fig:weights_mlp0_seqall}
% \end{figure}

\subsection{Population-level Representations}

\begin{figure}
    \includegraphics[width=1\textwidth]{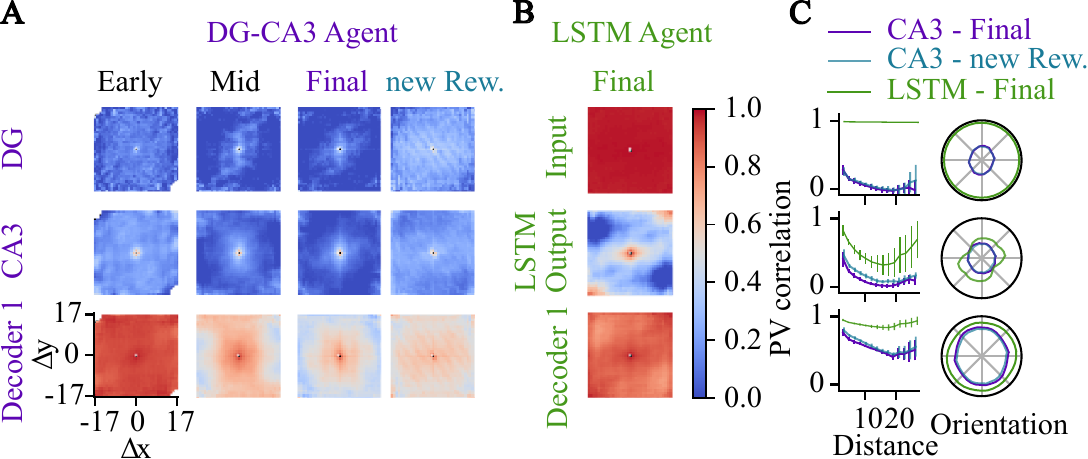}
    \caption{Kernel representation of distance. \textbf{A} Mean Pearson correlation (color code) of pairs of population activity vectors (PV) as a function of spatial displacement. Rows correspond to different layers (brain regions) of the network, columns to learning stages (epochs). \textbf{B} Same as \textbf{A}, but in dense LSTM agents. \textbf{C} PV correlation as a function of distance (left) and orientation (right).}
    \label{fig:correlation_kernel}
\end{figure}

We next examined population-level representations. We employ a novel method inspired by population vector (PV) correlation and representational (dis)similarity analysis~\citep{Kornblith2019Similarity,Kriegeskorte2021Neural}. It measures the mean population vector correlation in a pair of location bins  grouped by the spatial displacement $\Delta x,\Delta y$. This measure can be interpreted as a spatial kernel learned by the network (Fig.~\ref{fig:correlation_kernel}). An unbiased representation of spatial geometry would show a kernel function that is isotropic, and smoothly and monotonically depends on the distance between locations bins. Conversely, the representation could be restricted to specific location bin pairs, displacement along specific orientation, specific spatial frequency, or simple visual resemblance. Kernels from all layers (DG, CA3, and Decoder layers) exhibit some dependence on distance throughout learning. However, CA3 kernels are smoother than the DG kernels. Both DG and CA3 show lower correlation values compared to the decoder layer. Notably, the Decoder layer 1 develops the most pronounced spatial tuning (Fig.~\ref{fig:correlation_kernel}A), consistent with the single-unit selectivity observed in Fig.~\ref{fig:SI}. After the reward location was changed (“new”), kernels became less sharply defined, indicating that the representations are disrupted although behavior is adapted.
Most layers in the agent with LSTM core and dense input did not show spatial kernels with strong displacement dependency. Only the LSTM output units showed a gradually refined spatial kernel during learning but it is strongly non-isotropic (Fig.~\ref{fig:correlation_kernel}B; Fig.~\ref{fig:lstm-corr-kernel}). The dependence of representational similarity on distance and orientation is quantified in Fig.~\ref{fig:correlation_kernel}C.

Over training, a stable place code gradually emerges and remaps after learning the new reward location, indicated by the correlation of population vectors across training epochs (Fig.~\ref{fig:PV_corrs_epoch}). Interestingly, the similarity between fully trained networks (epoch 6) and the novel reward condition (“new”) is higher than between the naive (epoch 0) and trained (epoch 6) networks, suggesting that the agent acquires generalizable knowledge about the arena’s spatial layout.

% \subsection{Representations in the LSTM agent}
% %(Figs.~\ref{fig:lstm-corr-kernel}\ref{fig:lstm-dg-pfs}\ref{fig:lstm-occupancy}\ref{fig:lstm-pfs-core-all}\ref{fig:lstm-pf-corr-epochs})
% We performed the same analyses on the LSTM agent trained under dense input regime. The LSTM agent under dense sensory input has fewer converging trajectories before reaching the reward (Fig.~\ref{fig:lstm-occupancy}), likely due to the input being readily available. 
% It did not show orthogonalization of the learned input projection (Fig.~\ref{fig:lstm-dg-pfs}). 
% The spatial tuning of the LSTM units are disperse across the entire map, distinct from localized place-cell-like tunings (Fig.~\ref{fig:lstm-pfs-core-all}. 
% The spatial kernels in most layers do not show strong distance dependency. The LSTM core showed a gradually refined spatial kernel during learning but it is strongly non-isotropic (Fig.~\ref{fig:lstm-corr-kernel}). 
% These results demonstrate the distinction of learned strategies and representations in our hippocampus-inspired model and conventional LSTM-based model, even though the performance indicators are comparable.

\section{Discussion}

In this study, we presented a minimal model of the hippocampus that enables navigation in a vision-based virtual environment and reproduces key phenomena of spatial representation observed in the mammalian hippocampus.

The cornerstone of our model — the sequential connectivity in CA3 — is biologically plausible and inspired by established findings on theta sequence firing \citep{Dragoi2006-qq,Foster2007-wz,Yiu2022-kz}. By constraining CA3 recurrent connectivity as a dynamical reservoir and limiting training to input and output weights via reinforcement learning, we clearly isolate the effect of this theta sequence generator.

Agents equipped with the CA3 sequence module achieved superior navigation from sparse input (analogous to DG activity) compared to LSTMs of similar complexity or SSM-based cores. Crucially, this advantage was regime-dependent: under sparse input, CA3 dynamics and DG sparsification synergized to support robust navigation, whereas under dense input, conventional recurrent cores such as LSTMs performed better. Performance also degraded substantially when CA3 sequence length was ablated or shortened (Fig.~\ref{fig:training_DS}), highlighting the functional role of theta sequence generator.

Analysis of neural activities revealed clear parallels with experimental observations of place cell properties, including robust place field formation (Fig.~\ref{fig:Place_field_seq0}), population-level encoding of spatial distance (Fig.~\ref{fig:correlation_kernel}), progressive orthogonalization of DG outputs (Fig.~\ref{fig:Epoch_DG_entropy_SI}A), and dynamic remapping triggered by changes in reward locations (Fig.~\ref{fig:PV_corrs_epoch}~\ref{fig:CoM}), consistent with experimental findings \citep{leutgeb_pattern_2007,fenton_remapping_2024}.

\paragraph{Interpretation.}
The CA3 module expands sparse DG codes into a temporally smoothed canonical basis set, providing long-horizon history without the indiscriminate feature mixing of fully connected RNNs. This is especially beneficial under sparse input, where immediate sensory signals are limited and long-range context is critical for policy learning. The repetition parameter $R$ functions as a built-in prior of the temporal smoothness of latent states, thereby smoothing the CA3 spatial tuning via the agent’s movement, at the cost of imprecise timing of input memory.

At first sight, sparsifying input to a recurrent core seems purely detrimental—leaving the agent “blind” most of the time. Yet it also filters out noisy, non-informative cues, making each supra-threshold input more reliably tuned to restricted regions of space. Through sequence propagation and policy learning, a spatially smooth representations can then be stabilized. This mechanism is consistent with habitual trajectories: neuronal activity late in a sequence only remains informative if the policy converges onto consistent paths.

Under dense input, LSTMs perform better, likely because the abundant weak signals can be accumulated to inform space, reducing the need for long-horizon buffering of strong signals.
These intuitions are supported by comparative experiment where different levels of Gaussian noise were added to the pixel input to suppress weak signals. The DG+CA3 agent suffers less from increasing noise level than the LSTM agent (Fig.~\ref{fig:noise}; Tab.~\ref{tab:nosie}). 

% The CA3 module simply propagates the DG input, with repeat number R rigidly combining adjacent inputs together (as compared to other RNNs which could have various input response functions). It helps to preserve the timing information of the input, but requires the input to be highly informative. Thus CA3 benefits specifically from thresholding that improves the information per input at the cost of reducing total amount of inputs. 

Overall, these results suggest that different recurrent architectures are suited to different sensory regimes: sparse coding naturally complements sequential expansion, whereas dense input favors mixing-oriented recurrent dynamics. The agents' behaviors also diverged, resembling memory-driven navigation in our model versus more sensory-driven “visual search” navigation with LSTMs.

\paragraph{Biological and computational relevance.}
We proposed a parsimonious account of hippocampal theta sequences: they can be intrinsically maintained without requiring external input. While actual theta sequences in biological systems likely arise from multiple factors and vary across contexts~\citep{Chance2012-qq,Schlesiger2015-io,Yiu2023-bv,Ahmadi2025-lj}, our abstraction nonetheless reproduced hallmark hippocampal phenomena, including robust place fields (Fig.~\ref{fig:Place_field_seq0}), progressive DG orthogonalization (Fig.~\ref{fig:Epoch_DG_entropy_SI}A), distance-dependent population kernels (Fig.~\ref{fig:correlation_kernel}), remapping after goal changes (Fig.~\ref{fig:PV_corrs_epoch} and Fig.~\ref{fig:CoM}), and, by construction, theta sequences \citep{Foster2007-wz}. 
These representational effects were linked to performance and align with evidence on sparse DG activity and CA3 sequences \citep{leutgeb_pattern_2007,fenton_remapping_2024}. 

% --- EC discussion inserted here ---
Importantly, the visual encoder can be viewed as capturing the visually driven components of entorhinal cortex (EC) input to DG and CA3: the dorsal “where” and ventral “what” streams that project to medial and lateral EC \citep{wang2011gateways}. The depth sensor signal fed to Decoder can be seen as temporoamonic pathway from EC to CA1. To reflect the multi-modal nature of EC, additional modalities could be incorporated as input features.
% --- end EC discussion ---

Beyond explaining brain phenomena, our brain-inspired minimal model also proved useful as a module in competitive deep RL agents. The CA3 shift register can be viewed as a sparsely active reservoir that generates finite-length temporal bases, contrasting with the rotational modes of Legendre SSMs and the fading modes of Laguerre SSMs. Its structure resonates with recent shift–diagonal architectures \citep{fu2022hungry}, but tuned to sparse sensory regimes and navigation tasks. 
Thus, the model not only explains how theta sequences and place cells form but also provides a normative account of the computational effectiveness of this mechanism for navigation.

%LSTM-< no place fields
The comparison between our hippocampus-inspired model and an LSTM agent with dense input demonstrate the distinction of learned strategies and representations, even though the performance indicators are comparable (Figs.~\ref{fig:lstm-corr-kernel}\ref{fig:lstm-dg-pfs}\ref{fig:lstm-occupancy}\ref{fig:lstm-pfs-core-all}\ref{fig:lstm-pf-corr-epochs}).

\paragraph{Biological Predictions.}
Our model predicts that larger environments or sparser inputs require longer sequences for successful navigation, consistent with developmental adaptability of sequence length \citep{wikenheiser2015hippocampal,Farooq2019-ty}. More broadly, it suggests that hippocampal spatial representations could rely largely on intrinsic sequence-generating circuitry, with experience primarily shaping feedforward and readout connections. It offers an explanation for how place cells can persist despite lesions of entorhinal cortex \citep{Hales2014-li,Schlesiger2015-io}. The parsimonious nature of our model also provides a unified mechanism how navigation can be built upon sequences, no matter how they arise and how they differ across species. This is particularly important in species without prominent theta oscillations, e.g. bats show hippocampal sequences locked to wingbeats \citep{forli2025bat_replay}.

\paragraph{Future Directions.}
Several extensions follow naturally. On the biological side, incorporating local plasticity rules into the DG–CA3 pathway or CA3 readouts would align the model more closely with known mechanisms while preserving the benefits of prewired dynamics. Interactions with path integration in medial EC could further clarify their complementary roles in spatial cognition. Moreover, theta sequences have also been observed in other brain areas, raising the possibility of hierarchical coordination across regions.

These ideas resonate with developments in machine learning: structured dynamics akin to SSMs and linear attention are increasingly used in modern sequence models, including LLMs \citep{fu2022hungry,katharopoulos2020transformers}, suggesting that hippocampal-like motifs may illuminate principles of efficient long-range computation. The simplicity of our model also makes it interpretable: sequentially connected CA3 units can be seen as structurally representing trajectories, which could in turn motivate algorithms that minimize internally measured trajectory length without relying on external reward.

On the ML side, combining sparse sequential reservoirs with learned SSMs may yield hybrid architectures capable of adapting flexibly across input regimes. More broadly, the bottom-up approach—focusing on circuit motifs rather than top-down expert informed objectives—suggests a path toward scaling to larger networks and more complex tasks.

\paragraph{RL for neuroscience.}
% Beyond ML implications, our framework illustrates how modern RL can serve as a testbed for neuroscience theories. By embedding a biologically inspired model within a reinforcement-learning agent, we can study brain computation beyond intermediate objectives such as sensory processing, cognition, or motor control, and consider the entire perception–action cycle—particularly useful for association cortices such as hippocampus and prefrontal cortex where representations reflect both the input and action planning.

% Furthermore, the behavioral and representational comparison between CA3 agent and LSTM agent showed a useful direction of research - qualitatively distinct strategies beyond performance benchmarking can provide useful insights for architectural designs.

Our framework showcases how modern RL can serve as a testbed for computational neuroscience. By embedding a biologically inspired hippocampal circuit within an end-to-end reinforcement-learning agent, the model enables us to study brain computation in the context of the full perception–action loop, rather than in isolated modules for sensory processing, cognition, or motor control. Our approach thereby relates to ongoing discussions in machine learning regarding the “Bitter Lesson”, which cautions against relying on handcrafted intermediate representations in favor of scalable end-to-end optimization. Importantly, our structural priors (sparseness, sequences) do not prescribe the representational solution; place fields, distance-dependent kernels, and remapping emerge through end-to-end learning. In this sense, physiological constraints act as an inductive bias that narrows the hypothesis space without sacrificing scalability, providing a middle ground between unconstrained black-box models and heavily engineered representational schemes.  This perspective -- imposing minimal structural constraints, such as sparse connectivity, rather than making apriori assumptions -- is particularly relevant for association cortices such as hippocampus and prefrontal cortex, where emerging representational patterns are shaped jointly by sensory inputs, movement statistics, and behavioral goals~\citep{lin2023}.

An important conceptual point concerns how our representations relate to existing theoretical frameworks. Our CA3 representations relate to both successor-representation (SR) and reward/value-based frameworks, but with important distinctions. As in SRs, CA3 activity is policy dependent and shows a consistent temporal ordering: because the CA3 module propagates an intrinsic sequence, each unit predicts its downstream neighbors. This produces SR-like anticipatory structure, yet it arises solely from the fixed physiology-inspired architecture—CA3 does not learn a predictive map and receives no TD updates or discounted occupancy signal.

At the same time, unlike many previous RL studies of hippocampus \citep{kumar2022actorcritic,leibold_model_2020}, that takes allocentric input and analyze how policy shapes and utilizes them, our contribution is complementary: we show how localized, Gaussian-like place tuning can emerge from egocentric observations when combined with sparse DG input and intrinsic CA3 sequences. Our work also distinguishes itself from many other work that takes egocentric input but where the spatial similarity structure is already reflected in the input (e.g. small room where each wall has a different color; \citet{VijayabaskaranC22cheng_navigation,raju_space_2024}). Our environment is large maze, with uniform looking obstacles scattered in the arena, accounting for a more realistic discrepancy between visual representation and spatial representation.

The behavioral and representational divergence between the CA3-agent and the LSTM-agent illustrates a promising direction for future comparative work. Differences in qualitative strategies—not only performance—can reveal how architectural constraints shape navigation, representational geometry, and generalization, providing insights that are difficult to obtain from performance benchmarks alone.

%In a similar spirit, our model does not prescribe spatial maps or predictive objectives; instead, sparse DG input and fixed sequential CA3 dynamics constrain the hypothesis space while leaving representational structure to be shaped by end-to-end reinforcement learning.

\paragraph{Conclusion}
A minimal, sparsely driven sequence generator in actor-critic agent not only supports successful navigation but also gives rise to hippocampus-like spatial representations.

\section{Acknowledgements}
This work was funded by the German federal ministry for research, technology and space travel (BMFTR) under grant number 01GQ2510 and the BrainLinks-BrainTools Center. 

\bibliography{Hipposlam}

\begin{thebibliography}{50}
\providecommand{\natexlab}[1]{#1}
\providecommand{\url}[1]{\texttt{#1}}
\expandafter\ifx\csname urlstyle\endcsname\relax
  \providecommand{\doi}[1]{doi: #1}\else
  \providecommand{\doi}{doi: \begingroup \urlstyle{rm}\Url}\fi

\bibitem[Ahmadi et~al.(2025)Ahmadi, Sasaki, Sabariego, Leibold, Leutgeb, and Leutgeb]{Ahmadi2025-lj}
Siavash Ahmadi, Takuya Sasaki, Marta Sabariego, Christian Leibold, Stefan Leutgeb, and Jill~K Leutgeb.
\newblock Distinct roles of dentate gyrus and medial entorhinal cortex inputs for phase precession and temporal correlations in the hippocampal {CA3} area.
\newblock \emph{Nat. Commun.}, 16\penalty0 (1):\penalty0 13, January 2025.

\bibitem[Amaral et~al.(2007)Amaral, Scharfman, and Lavenex]{Amaral}
David~G Amaral, Helen~E Scharfman, and Pierre Lavenex.
\newblock The dentate gyrus: fundamental neuroanatomical organization (dentate gyrus for dummies).
\newblock In \emph{The Dentate Gyrus: A Comprehensive Guide to Structure, Function, and Clinical Implications}, Progress in brain research, pp.\  3--790. Elsevier, 2007.

\bibitem[Aronowitz \& Nadel(2023)Aronowitz and Nadel]{aronowitz2023space}
Sara Aronowitz and Lynn Nadel.
\newblock \emph{Space, and not Time, Provides the Basic Structure of Memory}.
\newblock Oxford University Press, 2023.

\bibitem[Beattie et~al.(2016)Beattie, Leibo, Teplyashin, Ward, Wainwright, Küttler, Lefrancq, Green, Valdés, Sadik, Schrittwieser, Anderson, York, Cant, Cain, Bolton, Gaffney, King, Hassabis, Legg, and Petersen]{beattie2016deepmindlab}
Charles Beattie, Joel~Z. Leibo, Denis Teplyashin, Tom Ward, Marcus Wainwright, Heinrich Küttler, Andrew Lefrancq, Simon Green, Víctor Valdés, Amir Sadik, Julian Schrittwieser, Keith Anderson, Sarah York, Max Cant, Adam Cain, Adrian Bolton, Stephen Gaffney, Helen King, Demis Hassabis, Shane Legg, and Stig Petersen.
\newblock Deepmind lab, 2016.
\newblock URL \url{https://arxiv.org/abs/1612.03801}.

\bibitem[Buzs{\'a}ki \& Tingley(2018)Buzs{\'a}ki and Tingley]{buzsaki2018sequence_generator}
Gy{\"o}rgy Buzs{\'a}ki and David Tingley.
\newblock Space and time: the hippocampus as a sequence generator.
\newblock \emph{Trends in cognitive sciences}, 22\penalty0 (10):\penalty0 853--869, 2018.

\bibitem[Chance(2012)]{Chance2012-qq}
Frances~S Chance.
\newblock Hippocampal phase precession from dual input components.
\newblock \emph{J. Neurosci.}, 32\penalty0 (47):\penalty0 16693--703a, November 2012.

\bibitem[Dragoi \& Buzs{\'a}ki(2006)Dragoi and Buzs{\'a}ki]{Dragoi2006-qq}
George Dragoi and Gy{\"o}rgy Buzs{\'a}ki.
\newblock Temporal encoding of place sequences by hippocampal cell assemblies.
\newblock \emph{Neuron}, 50\penalty0 (1):\penalty0 145--157, April 2006.

\bibitem[Eichenbaum(2014)]{eichenbaum2014time}
Howard Eichenbaum.
\newblock Time cells in the hippocampus: a new dimension for mapping memories.
\newblock \emph{Nature Reviews Neuroscience}, 15\penalty0 (11):\penalty0 732--744, 2014.

\bibitem[Espeholt et~al.(2018)Espeholt, Soyer, Munos, Simonyan, Mnih, Ward, Doron, Firoiu, Harley, Dunning, et~al.]{espeholt2018impala}
Lasse Espeholt, Hubert Soyer, Remi Munos, Karen Simonyan, Vlad Mnih, Tom Ward, Yotam Doron, Vlad Firoiu, Tim Harley, Iain Dunning, et~al.
\newblock Impala: Scalable distributed deep-rl with importance weighted actor-learner architectures.
\newblock In \emph{International conference on machine learning}, pp.\  1407--1416. PMLR, 2018.

\bibitem[Farooq \& Dragoi(2019)Farooq and Dragoi]{Farooq2019-ty}
U~Farooq and G~Dragoi.
\newblock Emergence of preconfigured and plastic time-compressed sequences in early postnatal development.
\newblock \emph{Science}, 363\penalty0 (6423):\penalty0 168--173, January 2019.

\bibitem[Fenton(2024)]{fenton_remapping_2024}
André~A. Fenton.
\newblock Remapping revisited: how the hippocampus represents different spaces.
\newblock \emph{Nature Reviews Neuroscience}, 25\penalty0 (6):\penalty0 428--448, 2024.
\newblock ISSN 1471-0048.
\newblock \doi{10.1038/s41583-024-00817-x}.
\newblock URL \url{https://www.nature.com/articles/s41583-024-00817-x}.
\newblock Publisher: Nature Publishing Group.

\bibitem[Forli et~al.(2025)Forli, Fan, Qi, and Yartsev]{forli2025bat_replay}
Angelo Forli, Wudi Fan, Kevin~K Qi, and Michael~M Yartsev.
\newblock Replay and representation dynamics in the hippocampus of freely flying bats.
\newblock \emph{Nature}, 645\penalty0 (8082):\penalty0 974--980, 2025.

\bibitem[Foster \& Wilson(2007)Foster and Wilson]{Foster2007-wz}
David~J Foster and Matthew~A Wilson.
\newblock Hippocampal theta sequences.
\newblock \emph{Hippocampus}, 17\penalty0 (11):\penalty0 1093--1099, 2007.

\bibitem[Fu et~al.(2022)Fu, Dao, Saab, Thomas, Rudra, and R{\'e}]{fu2022hungry}
Daniel~Y Fu, Tri Dao, Khaled~K Saab, Armin~W Thomas, Atri Rudra, and Christopher R{\'e}.
\newblock Hungry hungry hippos: Towards language modeling with state space models.
\newblock \emph{arXiv preprint arXiv:2212.14052}, 2022.

\bibitem[Gibson et~al.(2013)Gibson, Butler, and Taube]{gibson2013head}
Brett Gibson, William~N Butler, and Jeffery~S Taube.
\newblock The head-direction signal is critical for navigation requiring a cognitive map but not for learning a spatial habit.
\newblock \emph{Current Biology}, 23\penalty0 (16):\penalty0 1536--1540, 2013.

\bibitem[Gu et~al.(2020)Gu, Dao, Ermon, Rudra, and R{\'e}]{gu2020hippo}
Albert Gu, Tri Dao, Stefano Ermon, Atri Rudra, and Christopher R{\'e}.
\newblock Hippo: Recurrent memory with optimal polynomial projections.
\newblock \emph{Advances in neural information processing systems}, 33:\penalty0 1474--1487, 2020.

\bibitem[Gu et~al.(2021)Gu, Goel, and R{\'e}]{gu2021efficiently}
Albert Gu, Karan Goel, and Christopher R{\'e}.
\newblock Efficiently modeling long sequences with structured state spaces.
\newblock \emph{arXiv preprint arXiv:2111.00396}, 2021.

\bibitem[Hales et~al.(2014)Hales, Schlesiger, Leutgeb, Squire, Leutgeb, and Clark]{Hales2014-li}
Jena~B Hales, Magdalene~I Schlesiger, Jill~K Leutgeb, Larry~R Squire, Stefan Leutgeb, and Robert~E Clark.
\newblock Medial entorhinal cortex lesions only partially disrupt hippocampal place cells and hippocampus-dependent place memory.
\newblock \emph{Cell Rep.}, 9\penalty0 (3):\penalty0 893--901, November 2014.

\bibitem[He et~al.(2015)He, Zhang, Ren, and Sun]{he2015deepresiduallearningimage_resnet}
Kaiming He, Xiangyu Zhang, Shaoqing Ren, and Jian Sun.
\newblock Deep residual learning for image recognition, 2015.
\newblock URL \url{https://arxiv.org/abs/1512.03385}.

\bibitem[Henze et~al.(2000)Henze, Urban, and Barrionuevo]{Henze2000-ag}
D~A Henze, N~N Urban, and G~Barrionuevo.
\newblock The multifarious hippocampal mossy fiber pathway: a review.
\newblock \emph{Neuroscience}, 98\penalty0 (3):\penalty0 407--427, 2000.

\bibitem[Hessel et~al.(2019)Hessel, Soyer, Espeholt, Czarnecki, Schmitt, and Van~Hasselt]{hessel2019multi_popart}
Matteo Hessel, Hubert Soyer, Lasse Espeholt, Wojciech Czarnecki, Simon Schmitt, and Hado Van~Hasselt.
\newblock Multi-task deep reinforcement learning with popart.
\newblock In \emph{Proceedings of the AAAI Conference on Artificial Intelligence}, volume~33, pp.\  3796--3803, 2019.

\bibitem[Jaeger \& Haas(2004)Jaeger and Haas]{Jaeger2004-le}
Herbert Jaeger and Harald Haas.
\newblock Harnessing nonlinearity: predicting chaotic systems and saving energy in wireless communication.
\newblock \emph{Science}, 304\penalty0 (5667):\penalty0 78--80, April 2004.

\bibitem[Jung \& McNaughton(1993)Jung and McNaughton]{Jung1993-ea}
M~W Jung and B~L McNaughton.
\newblock Spatial selectivity of unit activity in the hippocampal granular layer.
\newblock \emph{Hippocampus}, 3\penalty0 (2):\penalty0 165--182, April 1993.

\bibitem[Katharopoulos et~al.(2020)Katharopoulos, Vyas, Pappas, and Fleuret]{katharopoulos2020transformers}
Angelos Katharopoulos, Apoorv Vyas, Nikolaos Pappas, and Fran{\c{c}}ois Fleuret.
\newblock Transformers are rnns: Fast autoregressive transformers with linear attention.
\newblock In \emph{International conference on machine learning}, pp.\  5156--5165. PMLR, 2020.

\bibitem[Kay et~al.(2020)Kay, Chung, Sosa, Schor, Karlsson, Larkin, Liu, and Frank]{Kay2020-xz}
Kenneth Kay, Jason~E Chung, Marielena Sosa, Jonathan~S Schor, Mattias~P Karlsson, Margaret~C Larkin, Daniel~F Liu, and Loren~M Frank.
\newblock Constant sub-second cycling between representations of possible futures in the hippocampus.
\newblock \emph{Cell}, 180\penalty0 (3):\penalty0 552--567.e25, February 2020.

\bibitem[Kornblith et~al.(2019)Kornblith, Norouzi, Lee, and Hinton]{Kornblith2019Similarity}
Simon Kornblith, Mohammad Norouzi, Honglak Lee, and Geoffrey Hinton.
\newblock Similarity of neural network representations revisited.
\newblock In \emph{International conference on machine learning}, pp.\  3519--3529. PMLR, 2019.
\newblock ISBN 2640-3498.

\bibitem[Kriegeskorte \& Wei(2021)Kriegeskorte and Wei]{Kriegeskorte2021Neural}
Nikolaus Kriegeskorte and Xue-Xin Wei.
\newblock Neural tuning and representational geometry.
\newblock \emph{Nature Reviews Neuroscience}, 22\penalty0 (11):\penalty0 703--718, 11 2021.
\newblock ISSN 1471-0048.
\newblock \doi{10.1038/s41583-021-00502-3}.
\newblock number: 11 publisher: Nature Publishing Group.

\bibitem[Kuipers(2000)]{kuipers2000spatial_hierarchy}
Benjamin Kuipers.
\newblock The spatial semantic hierarchy.
\newblock \emph{Artificial intelligence}, 119\penalty0 (1-2):\penalty0 191--233, 2000.

\bibitem[Kumar et~al.(2022)Kumar, Tan, Libedinsky, Yen, and Tan]{kumar2022actorcritic}
M~Ganesh Kumar, Cheston Tan, Camilo Libedinsky, Shih-Cheng Yen, and Andrew Y~Y Tan.
\newblock A nonlinear hidden layer enables actor–critic agents to learn multiple paired association navigation.
\newblock \emph{Cerebral Cortex}, 32\penalty0 (18):\penalty0 3917--3936, 01 2022.
\newblock ISSN 1047-3211.
\newblock \doi{10.1093/cercor/bhab456}.
\newblock URL \url{https://doi.org/10.1093/cercor/bhab456}.

\bibitem[Leibold(2020)]{leibold_model_2020}
Christian Leibold.
\newblock A model for navigation in unknown environments based on a reservoir of hippocampal sequences.
\newblock \emph{Neural Networks}, 124:\penalty0 328--342, 2020.
\newblock ISSN 08936080.
\newblock \doi{10.1016/j.neunet.2020.01.014}.
\newblock URL \url{https://linkinghub.elsevier.com/retrieve/pii/S0893608020300162}.

\bibitem[Leibold(2022)]{leibold2022neuralkernel}
Christian Leibold.
\newblock Neural kernels for recursive support vector regression as a model for episodic memory.
\newblock \emph{Biological Cybernetics}, 116\penalty0 (3):\penalty0 377--386, 2022.

\bibitem[Leutgeb \& Leutgeb(2007)Leutgeb and Leutgeb]{leutgeb_pattern_2007}
Stefan Leutgeb and Jill~K. Leutgeb.
\newblock Pattern separation, pattern completion, and new neuronal codes within a continuous {CA}3 map.
\newblock \emph{Learning \& Memory}, 14\penalty0 (11):\penalty0 745--757, 2007.
\newblock ISSN 1072-0502, 1549-5485.
\newblock \doi{10.1101/lm.703907}.
\newblock URL \url{http://learnmem.cshlp.org/content/14/11/745}.
\newblock Company: Cold Spring Harbor Laboratory Press Distributor: Cold Spring Harbor Laboratory Press Institution: Cold Spring Harbor Laboratory Press Label: Cold Spring Harbor Laboratory Press Publisher: Cold Spring Harbor Lab.

\bibitem[Lin et~al.(2023)Lin, Nieder, and Jacob]{lin2023}
Xiao-Xiong Lin, Andreas Nieder, and Simon~N. Jacob.
\newblock The neuronal implementation of representational geometry in primate prefrontal cortex.
\newblock \emph{Science Advances}, 9\penalty0 (50):\penalty0 eadh8685, 2023.
\newblock \doi{10.1126/sciadv.adh8685}.
\newblock URL \url{https://www.science.org/doi/abs/10.1126/sciadv.adh8685}.

\bibitem[Mattar \& Daw(2018)Mattar and Daw]{Mattar2018-px}
Marcelo~G Mattar and Nathaniel~D Daw.
\newblock Prioritized memory access explains planning and hippocampal replay.
\newblock \emph{Nat. Neurosci.}, 21\penalty0 (11):\penalty0 1609--1617, November 2018.

\bibitem[O'Keefe \& Dostrovsky(1971)O'Keefe and Dostrovsky]{OKeefe1971-mr}
J~O'Keefe and J~Dostrovsky.
\newblock The hippocampus as a spatial map. preliminary evidence from unit activity in the freely-moving rat.
\newblock \emph{Brain Res.}, 34\penalty0 (1):\penalty0 171--175, November 1971.

\bibitem[Parra-Barrero et~al.(2021)Parra-Barrero, Diba, and Cheng]{parra2021sequence_speed_width}
Eloy Parra-Barrero, Kamran Diba, and Sen Cheng.
\newblock Neuronal sequences during theta rely on behavior-dependent spatial maps.
\newblock \emph{Elife}, 10:\penalty0 e70296, 2021.

\bibitem[Petrenko et~al.(2020)Petrenko, Huang, Kumar, Sukhatme, and Koltun]{petrenko2020sf}
Aleksei Petrenko, Zhehui Huang, Tushar Kumar, Gaurav~S. Sukhatme, and Vladlen Koltun.
\newblock Sample factory: Egocentric 3d control from pixels at 100000 {FPS} with asynchronous reinforcement learning.
\newblock In \emph{Proceedings of the 37th International Conference on Machine Learning, {ICML} 2020, 13-18 July 2020, Virtual Event}, volume 119 of \emph{Proceedings of Machine Learning Research}, pp.\  7652--7662. {PMLR}, 2020.
\newblock URL \url{http://proceedings.mlr.press/v119/petrenko20a.html}.

\bibitem[Raju et~al.(2024)Raju, Guntupalli, Zhou, Wendelken, Lázaro-Gredilla, and George]{raju_space_2024}
Rajkumar~Vasudeva Raju, J.~Swaroop Guntupalli, Guangyao Zhou, Carter Wendelken, Miguel Lázaro-Gredilla, and Dileep George.
\newblock Space is a latent sequence: A theory of the hippocampus.
\newblock \emph{Science Advances}, 10\penalty0 (31):\penalty0 eadm8470, 2024.
\newblock ISSN 2375-2548.
\newblock \doi{10.1126/sciadv.adm8470}.
\newblock URL \url{https://www.science.org/doi/10.1126/sciadv.adm8470}.

\bibitem[Schaeffer et~al.(2022)Schaeffer, Khona, and Fiete]{schaeffer2022nofreelunch}
Rylan Schaeffer, Mikail Khona, and Ila Fiete.
\newblock No free lunch from deep learning in neuroscience: A case study through models of the entorhinal-hippocampal circuit.
\newblock \emph{Advances in neural information processing systems}, 35:\penalty0 16052--16067, 2022.

\bibitem[Schlesiger et~al.(2015)Schlesiger, Cannova, Boublil, Hales, Mankin, Brandon, Leutgeb, Leibold, and Leutgeb]{Schlesiger2015-io}
Magdalene~I Schlesiger, Christopher~C Cannova, Brittney~L Boublil, Jena~B Hales, Emily~A Mankin, Mark~P Brandon, Jill~K Leutgeb, Christian Leibold, and Stefan Leutgeb.
\newblock The medial entorhinal cortex is necessary for temporal organization of hippocampal neuronal activity.
\newblock \emph{Nat. Neurosci.}, 18\penalty0 (8):\penalty0 1123--1132, August 2015.

\bibitem[Schulman et~al.(2017)Schulman, Wolski, Dhariwal, Radford, and Klimov]{schulman2017ppo}
John Schulman, Filip Wolski, Prafulla Dhariwal, Alec Radford, and Oleg Klimov.
\newblock Proximal policy optimization algorithms.
\newblock \emph{arXiv preprint arXiv:1707.06347}, 2017.

\bibitem[Skaggs et~al.(1992)Skaggs, McNaughton, and Gothard]{NIPS1992_skaggs}
William Skaggs, Bruce McNaughton, and Katalin Gothard.
\newblock An information-theoretic approach to deciphering the hippocampal code.
\newblock In S.~Hanson, J.~Cowan, and C.~Giles (eds.), \emph{Advances in Neural Information Processing Systems}, volume~5. Morgan-Kaufmann, 1992.
\newblock URL \url{https://proceedings.neurips.cc/paper_files/paper/1992/file/5dd9db5e033da9c6fb5ba83c7a7ebea9-Paper.pdf}.

\bibitem[Stachenfeld et~al.(2017)Stachenfeld, Botvinick, and Gershman]{stachenfeld2017hippocampus}
Kimberly~L Stachenfeld, Matthew~M Botvinick, and Samuel~J Gershman.
\newblock The hippocampus as a predictive map.
\newblock \emph{Nature neuroscience}, 20\penalty0 (11):\penalty0 1643--1653, 2017.

\bibitem[Tolman(1948)]{tolman1948cognitive}
Edward~C Tolman.
\newblock Cognitive maps in rats and men.
\newblock \emph{Psychological review}, 55\penalty0 (4):\penalty0 189, 1948.

\bibitem[Vijayabaskaran \& Cheng(2022)Vijayabaskaran and Cheng]{VijayabaskaranC22cheng_navigation}
Sandhiya Vijayabaskaran and Sen Cheng.
\newblock Navigation task and action space drive the emergence of egocentric and allocentric spatial representations.
\newblock \emph{PLoS Comput. Biol.}, 18\penalty0 (10):\penalty0 1010320, 2022.
\newblock \doi{10.1371/JOURNAL.PCBI.1010320}.
\newblock URL \url{https://doi.org/10.1371/journal.pcbi.1010320}.

\bibitem[Wang et~al.(2011)Wang, Gao, and Burkhalter]{wang2011gateways}
Quanxin Wang, Enquan Gao, and Andreas Burkhalter.
\newblock Gateways of ventral and dorsal streams in mouse visual cortex.
\newblock \emph{Journal of Neuroscience}, 31\penalty0 (5):\penalty0 1905--1918, 2011.

\bibitem[Wang et~al.(2024)Wang, Di~Tullio, Rooke, and Balasubramanian]{wang_time_2024}
Zhaoze Wang, Ronald~W. Di~Tullio, Spencer Rooke, and Vijay Balasubramanian.
\newblock Time makes space: Emergence of place fields in networks encoding temporally continuous sensory experiences.
\newblock \emph{Advances in Neural Information Processing Systems}, 37:\penalty0 37836--37864, 2024.
\newblock URL \url{https://proceedings.neurips.cc/paper_files/paper/2024/hash/42aae0e655c77d93edad9171ad9f4717-Abstract-Conference.html}.

\bibitem[Wikenheiser \& Redish(2015)Wikenheiser and Redish]{wikenheiser2015hippocampal}
Andrew~M Wikenheiser and A~David Redish.
\newblock Hippocampal theta sequences reflect current goals.
\newblock \emph{Nature neuroscience}, 18\penalty0 (2):\penalty0 289--294, 2015.

\bibitem[Yiu \& Leibold(2023)Yiu and Leibold]{Yiu2023-bv}
Yuk-Hoi Yiu and Christian Leibold.
\newblock A theory of hippocampal theta correlations accounting for extrinsic and intrinsic sequences.
\newblock \emph{Elife}, 12\penalty0 (RP86837), October 2023.

\bibitem[Yiu et~al.(2022)Yiu, Leutgeb, and Leibold]{Yiu2022-kz}
Yuk-Hoi Yiu, Jill~K Leutgeb, and Christian Leibold.
\newblock Directional tuning of phase precession properties in the hippocampus.
\newblock \emph{J. Neurosci.}, 42\penalty0 (11):\penalty0 2282--2297, March 2022.

\end{thebibliography}
\bibliographystyle{iclr2026_conference}
\appendix
\renewcommand{\thefigure}{A\arabic{figure}}
\renewcommand{\thetable}{A\arabic{table}}

% \section{A nice appendix}
\setcounter{figure}{0}   
\setcounter{table}{0}   
\section*{Appendix}
\section{Limitations}

\paragraph{Physiological Limitations}
Our model has fixed sequential connectivity in CA3, ignoring dynamic processes such as synaptic turnover, plasticity, and sequence length adaptations observed biologically. This simplification may limit the biological realism and generalizability of the model to adaptive neural processes. Additionally, the absence of direct entorhinal cortex inputs means we have not modeled potential interactions between hippocampal sequences and detailed sensory-motor driven activity.

\paragraph{Conceptual Assumptions}
We assume sequential activity in CA3 arises exclusively from connectivity rather than sequential inputs, potentially oversimplifying hippocampal dynamics. This assumption excludes possible interactions between external sequential signals and intrinsic CA3 network dynamics, which could affect real-world predictive accuracy.

\paragraph{Reinforcement Learning Methodological Limitations}
Our gradient-based reinforcement learning approach lacks clear biological analogs, as organisms likely utilize local Hebbian synaptic updates rather than global gradient propagation. Furthermore, the biological interpretation of the multilayer perceptron decoder remains unclear, posing challenges to direct biological interpretations of our results. Additionally, our use of asynchronous, batch-based training may not accurately reflect the temporal continuity or real-time single-organism learning dynamics seen biologically, thus potentially limiting ecological validity.

\paragraph{Technical Limitations}
Our results depend on the environment specifications and hyperparameter settings. We chose a reasonably difficult environment setting where the difference between architectures can be highlighted. It does not imply the effect can generalize to all environments. We tested the agent with L=64 and R=8 on 5 other randomly generated maps, it could converge at a similar speed with the map shown in other results (Fig.~\ref{fig:training_newmaps}). We also tested different frame-skip number, larger frame-skip leads to faster learning, and the effect of sequence length on performance is weaker (Fig.~\ref{fig:training_techgrid:both}). This is probably due to the fact that our environment has sparse obstacles and that larger frame-skip leads to effectively longer sequence and memory. Most results in our report use 4 policies for population based training and 8 environments per worker. These parameters also change the dynamics of learning. having more policies made agents with L=64 learn slower and L=8 learn slower. Further systematic investigation of these factors is required to establish broader applicability.

Taken together, these limitations highlight important areas for further experimental and theoretical investigation.

\section{Implementation Details}
\subsection{Environment}
\begin{table}[h]
\centering
\begin{tabular}{lrrrrrrr}
\toprule
Description & Look (dx) & Pitch (dy) & Strafe & Forward & Fire & Jump & Crouch \\
\midrule
Forward              & 0   & 0 & 0  & 1 & 0 & 0 & 0 \\
Strafe Left          & 0   & 0 & -1 & 0 & 0 & 0 & 0 \\
Strafe Right         & 0   & 0 & 1  & 0 & 0 & 0 & 0 \\
Look Left + Forward  & -20 & 0 & 0  & 1 & 0 & 0 & 0 \\
Look Right + Forward & 20  & 0 & 0  & 1 & 0 & 0 & 0 \\
\bottomrule
\end{tabular}
\caption{Reduced action set used in DeepMind Lab experiments. Each action is defined over the 7-dimensional control space \texttt{(look, pitch, strafe, forward, fire, jump, crouch)}.}
\label{tab:reduced_action_set}
\end{table}

Navigation task does not require complex actions, thus we reduced the action space used in IMPALA~\citep{espeholt2018impala}, to facilitate training speed (Table~\ref{tab:reduced_action_set}).

Since we are mainly interested in modeling the way-finding aspect of navigation~\citep{tolman1948cognitive,kuipers2000spatial_hierarchy}, the MLP also receives average depth of the pixels, down-sampled to 10 horizontal pixels %\(10\times1\), 
to aid motor control (going around walls and avoiding collisions). This procedure is supposed to mimic the sensory motor collision avoidance habits that do not depend on hippocampus. This depth information is directly routed to the decoder layer, so the representations in the recurrent core and its input are not directly influenced. 
% \paragraph{Training objective}  

\subsection{Multi-feature dynamics}\label{app:mf}
As a more direct illustration for the recurrent dynamics of the sequence network $X_{t+1} = A\, X_t + B\, u_t$, we, here, provide the matrices $A$ and be for the multi-feature case:
\begin{equation}
A =
\begin{bmatrix}
S&0&0&\dots&0\\
0&S&0&\dots&0\\
\vdots & &\ddots&&\vdots\\
0&\dots&&&S
\end{bmatrix}_{F\ell\times F\ell},
\qquad
B = \begin{bmatrix}
J&0&0&\dots&0\\
0&J&0&\dots&0\\
\vdots & &\ddots&&\vdots\\
0&\dots&&&J
\end{bmatrix}_{F\ell\times F}\ .
\end{equation}
with
\begin{equation}
S =
\begin{bmatrix}
0 & 0 & \dots & 0\\
1 & 0 & \dots & 0\\
 & \ddots & \ddots & \vdots\\
0 & \dots & 1 & 0
\end{bmatrix}_{\ell\times\ell},
\qquad
J = \bigl[\underbrace{1,1,\dots,1}_{R\text{ times}}\;,\underbrace{0,0,\dots,0}_{L-1\text{ times}}\bigr]^T \in \mathbb R^{\ell\times1}.
\end{equation}
Note that the rows in $A$ corresponding to the first time step of each feature only contain 0s, i.e., sequences can only be started from DG input and not from the dynamics itself. 

\subsection{Architecture}
\begin{table}[h]
\centering
\caption{Network architecture of the CA3 agent.}
\label{tab:hipposlam-arch}
\resizebox{\linewidth}{!}{
\begin{tabular}{ll}
\toprule
\textbf{Component} & \textbf{Details} \\
\midrule
\textbf{Encoder (ResNet)} & 3 conv blocks with residual layers, 16–32 channels, max pooling \\
\textbf{MLP head} & Linear (3456 $\to$ 256), ReLU \\
\textbf{DG projection} & Linear (256 $\to$ 16), BatchNorm (momentum 0.05, no affine), ReLU, intercept 2.43 \\
\textbf{CA3 (recurrent core)} & Fixed shift-register reservoir, size 16$\times$(64+8-1)=1136 \\
\textbf{Decoder} & MLP: Linear (1136 + 10 $\to$ 128), ReLU, Linear (128 $\to$ 128), ReLU \\
\textbf{Critic} & Linear (128 $\to$ 1) \\
\textbf{Actor} & Linear (128 $\to$ 5)  \\
\bottomrule
\end{tabular}}
\end{table}
CA3 agent has learnable parameters in DG projection and from CA3 to Decoder, the CA3 output in the full version has 1136*128=149504 parameters.
LSTM has learnable parameters $4*(m^2+mn)$ where m is hidden size and n is input and output size. Solving this gives a hidden size of roughly 137, which we used in the implementation.
SSM agents have matching state size with CA3 agent. They also have block-diagonal recurrent weights, where the weights in each block are obtained through the original implementation by \citet{gu2020hippo} with zero-order-hold discretizations.

\subsection{Actor Critic}
We optimize an actor–critic objective with advantage estimates as in 
\citet{espeholt2018impala}, consisting of a clipped policy loss, a value 
regression loss, and an entropy bonus:
\[
\mathcal{L}(\theta,\phi) =
  -\mathbb{E}_t\!\left[\min\!\bigl(r_t(\theta) A_t,\,
    \text{clip}(r_t(\theta),1-\epsilon,1+\epsilon)A_t\bigr)\right]
  + c_v\,\mathbb{E}_t[(R_t - V_\phi(s_t))^2]
  - c_e\,\mathbb{E}_t[\mathcal{H}(\pi_\theta(\cdot\mid s_t))],
\]
where \( r_t(\theta) = \pi_\theta(a_t\mid s_t)/\pi_{\theta_{\text{old}}}(a_t\mid s_t) \).  

Here \(R_t\) and the advantage estimates \(A_t\) are computed using 
V-trace returns \citep{espeholt2018impala}, which correct for 
off-policy updates arising in the asynchronous actor–learner setup 
of Sample Factory \citep{petrenko2020sf}. This formulation—known as 
Asynchronous PPO (APPO)—combines PPO’s clipped surrogate objective 
\citep{schulman2017ppo} with IMPALA’s V-trace corrections, enabling 
scalable training with many actors while maintaining stable policy 
updates.
\subsection{Training configuration}

We trained agents using \texttt{sample-factory} \citep{petrenko2020sf} with the following setup.

\paragraph{Environment.} 
frameskip 8, repeating action for 8 frames until getting the next observation. 
Each run used 32 workers $\times$ 8 envs/worker, with decorrelation up to 120s. 

\paragraph{Algorithm.} 
APPO \citep{espeholt2018impala}, $\gamma=0.99$, rollout length 64, recurrence 64, batch size 2048, 2 batches/epoch.  
Optimizer learning rate $2\!\times\!10^{-4}$.  

\paragraph{Architecture.} 
Visual encoder: pretrained ResNet on DMLab and the second layer of ResNet pretrained on ImageNet 
DG: batchnorm + ReLU, intercept 2.43, with 16 features.  
Decoder: 2 MLP layers of size 128.  

\paragraph{Population Based Training.} 
Enabled PBT with 4 policies, replacement gaps 0.05 (relative) and 0.2 (absolute), 
mutation start after 10M steps, period every 2M steps. 
Policy lag tolerance set to 35.  

\paragraph{Logging and checkpoints.} 
Training for 108k seconds, milestones every 5400s. 

\paragraph{Miscellaneous.} 
Seeds: [1111,2222,3333,4444,5555]. Device: CPU. 
Affinity pinning disabled.   
Inputs not normalized. 
Other parameters were default in sample-factory for DeepMind Lab.
% \section{Statistics}
\subsection{Skaggs’ Spatial Information Measure}

We quantify how much information a unit's activity conveys about the agent’s location in our discrete‐time simulation by a “bits per step” version of Skaggs’ spatial information~\citep{NIPS1992_skaggs}.  Let the environment be divided into \(N\) spatial bins, and define:

\begin{itemize}
  \item \(p_i\) — fraction of timesteps (steps) spent in bin \(i\) (occupancy probability),
  \item \(\lambda_i\) — mean activity rate in bin \(i\), measured in activity per step,
  \item \(\lambda = \sum_{i=1}^N p_i\,\lambda_i\) — overall activity rate (activity per step).
\end{itemize}

First, the information conveyed per timestep is
\[
  I_{\mathrm{step}}
  \;=\;
  \sum_{i=1}^{N}
    p_i\,
    \lambda_i
    \log_{2}\!\biggl(\frac{\lambda_i}{\lambda}\biggr)
  \quad\text{[bits/step].}
\]
This measures the average reduction in positional uncertainty (in bits) each simulation step provides.

% \paragraph{Bits per unit activity.}  
% Often one normalizes by the mean firing rate to get information per spike:
% \[
%   I_{\mathrm{spike}}
%   \;=\;
%     \frac{1}{\lambda}\,
%     I_{\mathrm{step}}
%   \;=\;
%     \sum_{i=1}^{N}
%       p_i \,
%       \frac{\lambda_i}{\lambda}
%       \log_{2}\!\biggl(\frac{\lambda_i}{\lambda}\biggr)
%   \quad\text{[bits/unit activity].}
% \]

Interpretation:
\begin{itemize}
  \item Uniform firing (\(\lambda_i=\lambda\) for all \(i\)) yields zero information (\(I_{\mathrm{step}}=I_{\mathrm{spike}}=0\)).  
  \item Elevated \(\lambda_i\) in particular bins gives positive contributions proportional to \(p_i\,\lambda_i\) (for bits/step) or \(p_i\,(\lambda_i/\lambda)\) (for bits/spike).  
  \item Bins visited rarely (small \(p_i\)) contribute less, guarding against over‐weighing seldom‐visited locations.
\end{itemize}
 
In our simulations, \(\lambda_i\) and \(\lambda\) are estimated from spatially binned sum of activity divided by the number of steps in each bin.  Bins with \(\lambda_i=0\) are omitted (treating \(0\cdot\log 0=0\)).
% , and a minimum occupancy threshold is enforced so that very sparsely visited bins do not spuriously inflate the information estimate.  
\newpage

\section{Supplementary Table}
\begin{table}[h]
\centering
\caption{The effect of parameter R at different running speed (frameskip). Performance is measured as the success rate at 150 million training frames. This is a summary of Fig.~\ref{fig:Hippo_R}.}
\label{tab:Hippo_R}
% \begin{tabular}{lcccccc}
% \toprule
% Frameskip & R=1 & R=4 & R=8 & R=12 & R=16 & R=32 \\
% \midrule
% 8  & 0.79$\pm$0.10 & 0.87$\pm$0.10 & 0.86$\pm$0.07 & 0.89$\pm$0.09 & 0.88$\pm$0.04 & 0.81$\pm$0.10 \\
% 4  & 0.79$\pm$0.15 & 0.85$\pm$0.13 & 0.87$\pm$0.11 & 0.90$\pm$0.12 & 0.90$\pm$0.11 & 0.76$\pm$0.23 \\
% \bottomrule
% \end{tabular}
\small
\begin{tabular}{lcccccc}
\toprule
Frameskip & R=1 & R=4 & R=8 & R=12 & R=16 & R=32 \\
\midrule
8  & 0.788$\pm$0.097 & 0.866$\pm$0.098 & 0.857$\pm$0.068 & 0.886$\pm$0.089 & 0.880$\pm$0.036 & 0.806$\pm$0.097 \\
4  & 0.785$\pm$0.153 & 0.848$\pm$0.126 & 0.874$\pm$0.106 & 0.902$\pm$0.116 & 0.901$\pm$0.114 & 0.758$\pm$0.226 \\
\bottomrule
\end{tabular}
\end{table}

\begin{table}[h]
\centering
\caption{The performance at different noise level. Performance is measured as the success rate at 100 million training frames. This is a summary of Fig.~\ref{fig:noise}.}
\label{tab:nosie}
\begin{tabular}{lccccc}
\toprule
Model $\backslash$ Noise & 0 & 10 & 20 & 40 & 80 \\
\midrule
DG+CA3      & 0.693 $\pm$ 0.137 & 0.492 $\pm$ 0.167 & 0.539 $\pm$ 0.108 & 0.685 $\pm$ 0.186 & 0.477 $\pm$ 0.072 \\
Dense LSTM  & 0.638 $\pm$ 0.063 & 0.707 $\pm$ 0.178 & 0.495 $\pm$ 0.010 & 0.449 $\pm$ 0.064 & 0.373 $\pm$ 0.008 \\
\bottomrule
\end{tabular}
\end{table}
\section{Supplementary Figures}
%TODOTODO: write better captions, more detailed descriptions

% \subsection{Occupancy}
% \begin{figure}
%     \centering
%     \includegraphics[width=1\linewidth]{Neuripsfigs/occu_epoch_vec.pdf}
%     \caption{Occupancy rate map throughout training. The agent's trajectory becomes more concentrated. }
%     \label{fig:occu_epoch}
% \end{figure}

%\subsection{Decoder Place Fields}

\begin{figure}[bht!]
    \centering
    \includegraphics[width=1\linewidth]{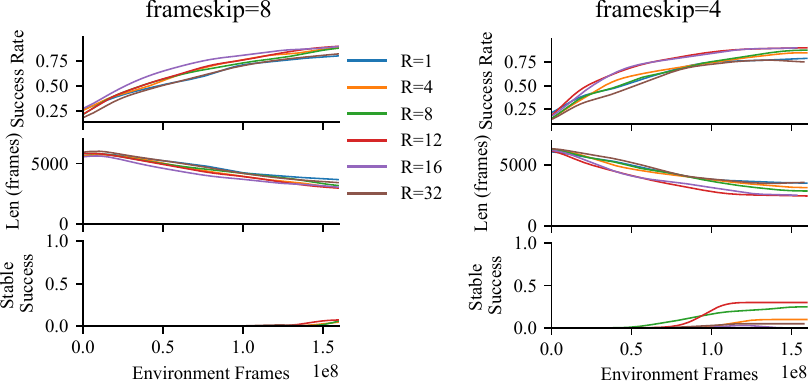}
    \caption{\textbf{Effect of parameter R with different running speed.} Training performance of CA3 agent with L=64 different R. Random seeds: [1111, 2222, 3333, 4444, 5555].}
    \label{fig:Hippo_R}
\end{figure}

\begin{figure}[bht!]
    \centering
    \includegraphics[width=0.4\linewidth]{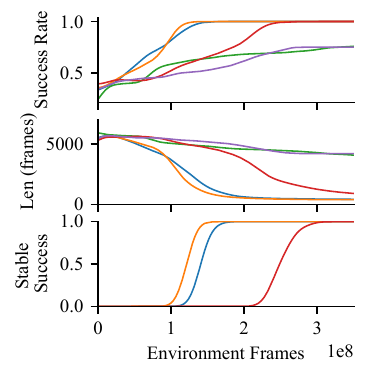}
    \caption{\textbf{Robustness of visual encoder.} Training performance of CA3 agent with the ResNet layer 2 output as visual encoder. Random seeds: [1111, 2222, 3333, 4444, 5555].}
    \label{fig:training_layer2}
\end{figure}

\begin{figure}
    \centering
    \includegraphics[width=0.5\linewidth]{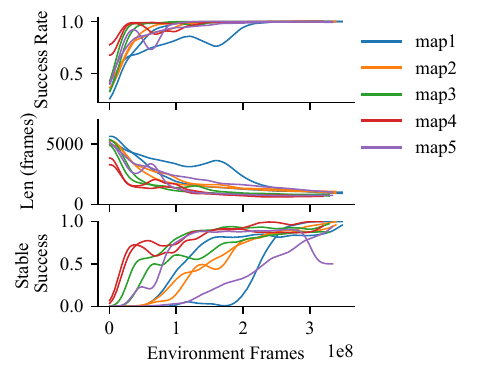}
    \caption{Training performance on 5 other randomly generated maps. Lines with the same color are results from random seeds [0, 42]}
    \label{fig:training_newmaps}
\end{figure}
\begin{figure}
    \centering
    \includegraphics[width=1\linewidth]{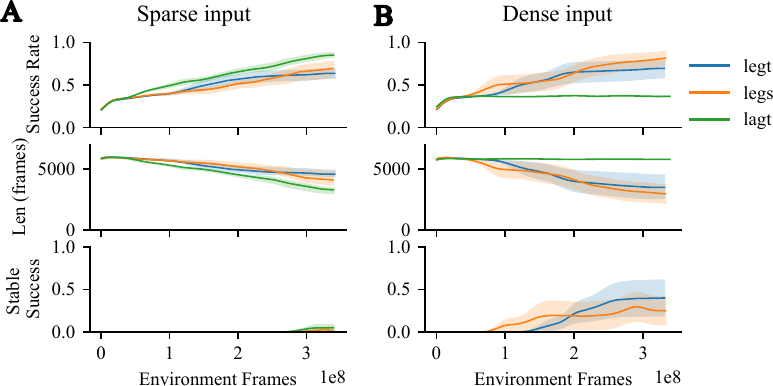}
    \caption{Training performance of agents with fixed SSM cores described by \citet{gu2020hippo}. legt: Legendre bases with fixed memory horizon. legs: Legendre bases with infinite memory from the beginning of an episode. lagt: Laguerre bases, i.e. memory with decay. All three SSMs were implemented with the same state size as our CA3 module, each input feature is expanded into 64+8-1=71 states. The SSMs were fixed, with three additional learnable parameter controlling the input, recurrent scales and timestep length for discretizations, as a common practice described by \citet{gu2021efficiently}.}
    \label{fig:training_SSM}
\end{figure}

\begin{figure}[bht!]
    \centering
    \includegraphics[width=1\linewidth]{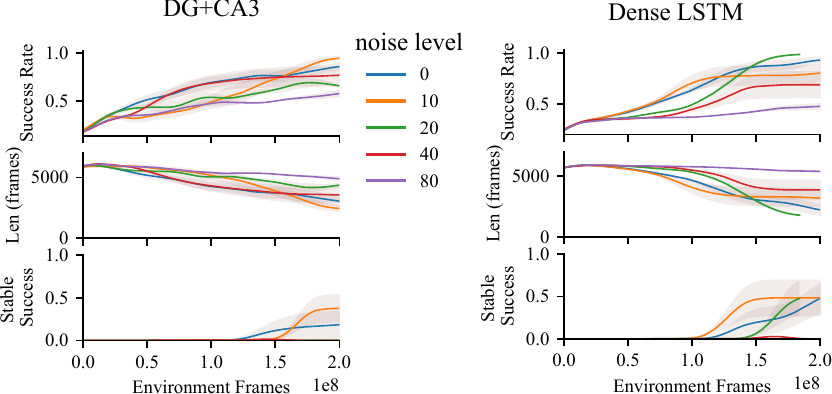}
    \caption{\textbf{Training performance of agents under different noise levels.} Independent Gaussian noise of different sigmas was added to the input image pixels (intensity 0-255). Random seeds: [1111, 2222, 3333, 4444, 5555].}
    \label{fig:noise}
\end{figure}

\begin{figure}[bht!]
    \centering
    \includegraphics[width=1\linewidth]{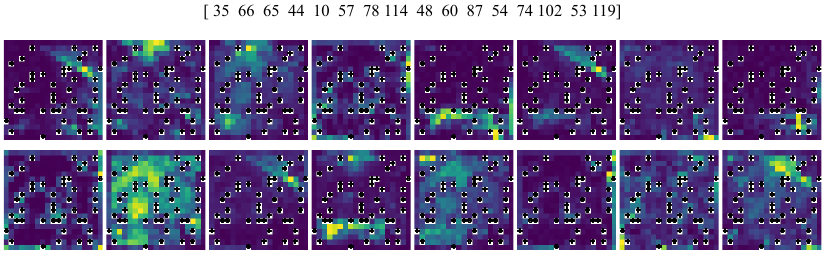}
    \caption{Rate maps from decoder layer 1 in the last training epoch. Showing the random 16 units out of the 50 units with the best SI. Unit id displayed on top.}
    \label{fig:Place_field_MLP0_rand}
\end{figure}

\begin{figure}[bht!]
    \centering
    \includegraphics[width=1\linewidth]{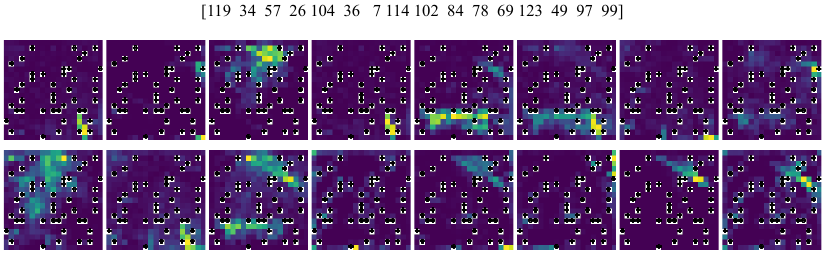}
    \caption{Rate maps from decoder layer 2 in the last training epoch. Showing the random 16 units out of the 50 units with the best SI. Unit id displayed on top.}
    \label{fig:Place_field_MLP2_rand}
\end{figure}

\begin{figure}
    % \hspace*{-0.2\textwidth}
    \includegraphics[width=0.85\textwidth]{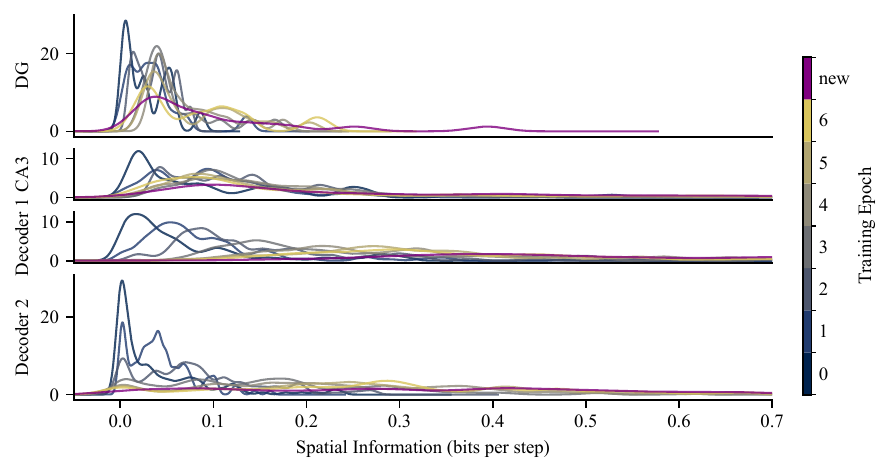}
    \caption{Distribution of SI rates for different layers (brain regions rows) over the progress of learning (colors) and change of reward site (purple). Y axes: unit fraction density.}
    \label{fig:SI}
\end{figure}

\begin{figure}
    \centering
    \includegraphics[width=0.8\linewidth]{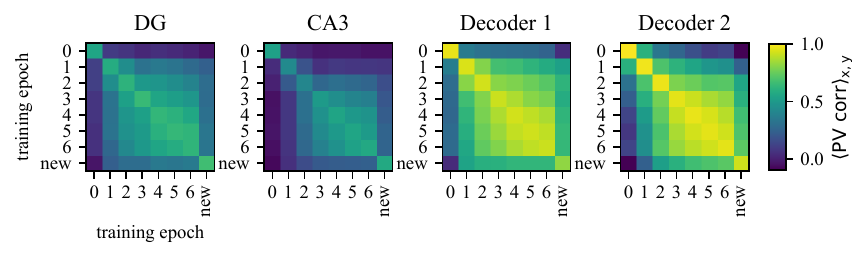}
    \caption{PV correlations across training. Although performance saturates by epochs 5–6, representational changes continue. The last epoch shows the effect of learning a new goal.}
    \label{fig:PV_corrs_epoch}
\end{figure}

\begin{figure}
    \centering
    \includegraphics[width=0.7\linewidth]{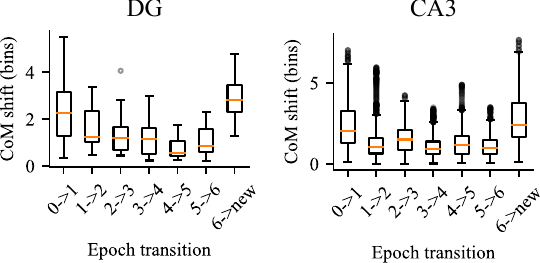}
    \caption{Convergence and Remapping. The center of mass (CoM) shift of place fields of individual units from epoch  to epoch reveals higher stability in DG and remapping for a new goal.}
    \label{fig:CoM}
\end{figure}

% \subsection{Representations in LSTM agents}
\begin{figure}[t]
  \centering
  \includegraphics[width=\linewidth]{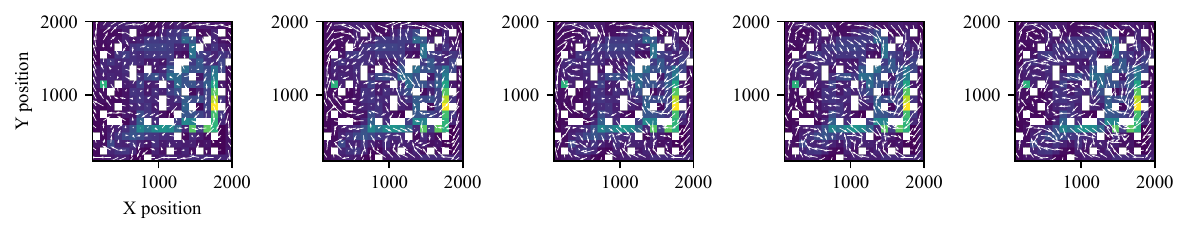}
  \caption{Occupancy map of LSTM agent's trajectories across training.}
  \label{fig:lstm-occupancy}
\end{figure}

\begin{figure}[t]
  \centering
  \includegraphics[width=\linewidth]{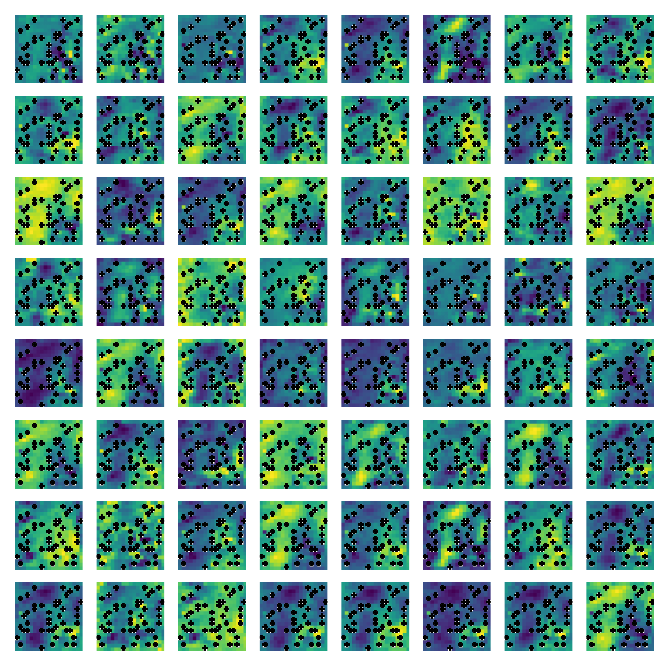}
  \caption{Place fields from the LSTM core (randomly selected units). Each panel shows the normalized spatial activity of one unit.}
  \label{fig:lstm-pfs-core-all}
\end{figure}

\begin{figure}[t]
  \centering
  \includegraphics[width=\linewidth]{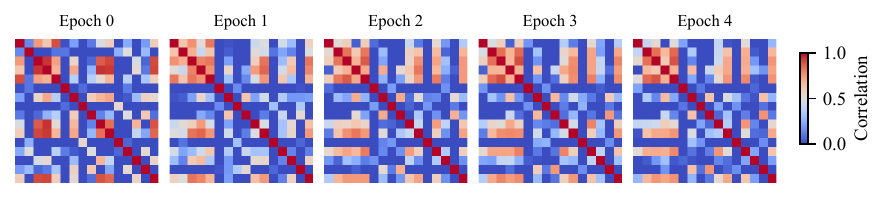}
  \caption{Population vector correlation across epochs. Color indicates Pearson correlation; rows show epochs 0–4.}
  \label{fig:lstm-pf-corr-epochs}
\end{figure}

\begin{figure}[t]
  \centering
  \includegraphics[width=\linewidth]{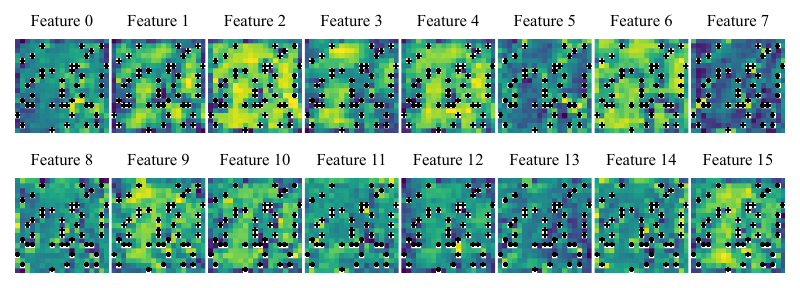}
  \caption{Learned dense input features in LSTM agent. Each tile shows the spatial receptive field of one feature.}
  \label{fig:lstm-dg-pfs}
\end{figure}

\begin{figure}[t]
  \centering
  \includegraphics[width=\linewidth]{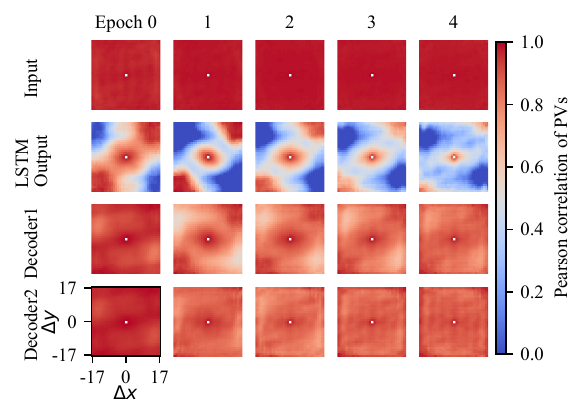}
  \caption{Displacement-dependent Pearson correlation kernel of population vectors. Axes show spatial shifts $(\Delta x,\Delta y)$; values are averaged correlations.}
  \label{fig:lstm-corr-kernel}
\end{figure}

\begin{figure}[bht!]
  \centering
  \begin{subfigure}[b]{0.48\linewidth}
    \centering
    \includegraphics[width=\linewidth]{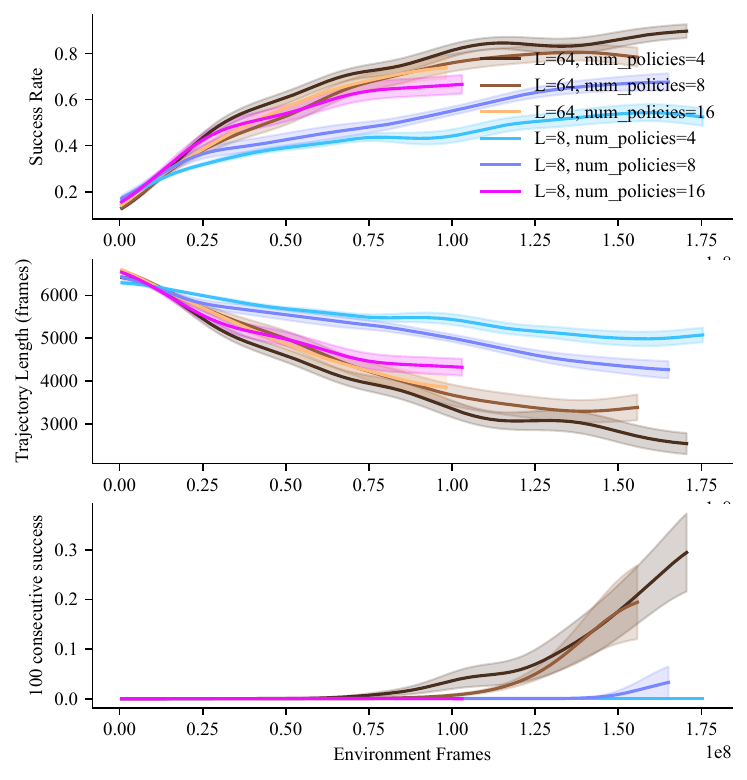}
    % \caption{Caption for the first panel.}
    \label{fig:techgrid:a}
  \end{subfigure}\hfill
  \begin{subfigure}[b]{0.48\linewidth}
    \centering
    \includegraphics[width=\linewidth]{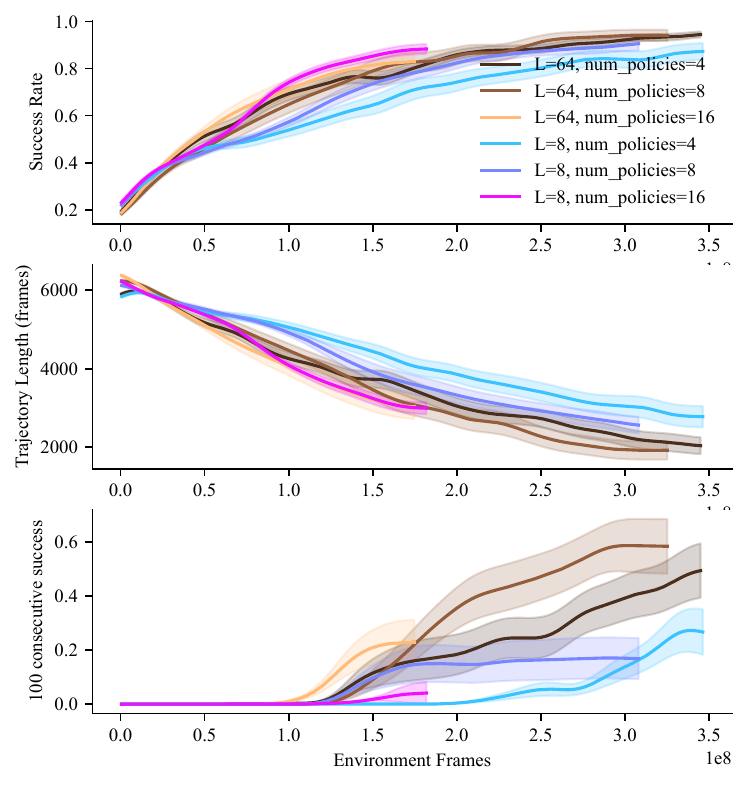}
    % \caption{Caption for the second panel.}
    \label{fig:techgrid:b}
  \end{subfigure}
  \caption{Training performance with different environment action repeats and number of policies in population based training. Left: environment frame skip / action repeat = 8. Right: environment frame skip / action repeat = 4. The frame skip controls the fine-graininess of actions, which is not required for the current navigation task. On the other hand, larger frame skip effectively make sequences propagate over longer traversals. }
  \label{fig:training_techgrid:both}
\end{figure}
\newpage
% \section{Implementation Notes}

% See supplement.

%%%%%%%%%%%%%%%%%%%%%%%%%%%%%%%%%%%%%%%%%%%%%%%%%%%%%%%%%%%%

\end{document}